\documentclass[prd,tightenlines,showpacs,nofootinbib,eqsecnum,amsfonts,amssymb,amsmath,twocolumn,floatfix]{revtex4-1}
\usepackage{graphicx}
\usepackage{amsmath}
\usepackage{amsfonts}   
\usepackage{amssymb}  
\usepackage{mathrsfs}
\usepackage{epsfig,bbm}
\usepackage{bm}
\usepackage{color}
\usepackage{enumerate}
\usepackage{longtable}

\newcommand{\obh}{$\Omega_{\mathrm{b}}{\cdot}h^2$}
\newcommand{\ob}{$\omega_{\mathrm{b}}$}

\newcommand{\deu}{${\rm D}$}
\newcommand{\tro}{$^3{\rm He}$}
\newcommand{\qua}{$^4{\rm He}$}

\newcommand{\sep}{$^{7}{\rm Li}$}

\newcommand{\hli}{$^4$He, D, $^3$He and $^{7}$Li}

\newcommand{\ddn}{d(d,n)$^3$He}
\newcommand{\ddp}{d(d,p)$^3$H}
\newcommand{\dpg}{d(p,$\gamma)^3$He}
\newcommand{\hag}{$^3$He($\alpha,\gamma)^7$Be}
\newcommand{\cago}{$^{12}$C($\alpha,\gamma)^{16}$O}

\newcommand{\sfac}{$S$--factor}
\newcommand{\nfac}{normalization factor}

\newcommand{\zaa}{Astron.~Astrophys.}

\newcommand{\zapj}{Astrophys.~J.}
\newcommand{\zapjl}{Astrophys.~J.~Lett.}
\newcommand{\zapjs}{Astrophys.~J.~Supp.}

\newcommand{\znat}{Nature}
\newcommand{\zepj}{Eur. Phys. J.}
\newcommand{\znp}{Nucl.~Phys.}
\newcommand{\znpa}{Nucl.~Phys.~A}
\newcommand{\zpl}{Phys.~Lett.}
\newcommand{\zpr}{Phys.~Rev.}
\newcommand{\zprd}{ Phys.~Rev. D}
\newcommand{\zprc}{ Phys.~Rev. C}
\newcommand{\zprl}{Phys.~Rev.~Lett.}

\newcommand{\znim}{Nucl.~Inst.~and~Meth.}
\newcommand{\zadndt}{At. Data Nucl. Data Tables}
\newcommand{\zmnras}{Mon. Not. R. Astron. Soc.}
\newcommand{\zzp}{Z.~Phys.}

\newcommand{\zjpg}{J. Phys. G}

\newcommand{\mnras}{MNRAS}
\newcommand{\zjcap}{Journal of Cosmology and Astroparticle Physics}

\newcommand{\etal}{{\it et al.}}

\begin{document}

\title{New reaction rates for improved primordial D/H calculation and the cosmic evolution of deuterium}

\author{Alain Coc}
 \email{coc@csnsm.in2p3.fr}
\affiliation {Centre de Sciences Nucl\'eaires et de Sciences de la
Mati\`ere  (CSNSM), Univ. Paris-Sud, CNRS/IN2P3, Universit\'e Paris--Saclay, 
B\^atiment 104, F--91405 Orsay Campus (France)} 

\author{Patrick Petitjean}
 \email{ppetitje@iap.fr}
\affiliation{Institut d'Astrophysique de Paris,
              UMR-7095 du CNRS, Universit\'e Pierre et Marie
              Curie,
              98 bis bd Arago, 75014 Paris (France),\\
              Sorbonne Universit\'es, Institut Lagrange de Paris, 98 bis bd Arago, 75014 Paris (France).}

\author{Jean-Philippe Uzan}
 \email{uzan@iap.fr}
 \affiliation{Institut d'Astrophysique de Paris,
              UMR-7095 du CNRS, Universit\'e Pierre et Marie
              Curie,
              98 bis bd Arago, 75014 Paris (France),\\
              Sorbonne Universit\'es, Institut Lagrange de Paris, 98 bis bd Arago, 75014 Paris (France).}

\author{Elisabeth Vangioni}
 \email{vangioni@iap.fr}
 \affiliation{Institut d'Astrophysique de Paris,
              UMR-7095 du CNRS, Universit\'e Pierre et Marie
              Curie,
              98 bis bd Arago, 75014 Paris (France),\\
              Sorbonne Universit\'es, Institut Lagrange de Paris, 98 bis bd Arago, 75014 Paris (France).}

\author{Pierre Descouvemont}
\email{pdesc@ulb.ac.be}
\affiliation{
Physique Nucl\'eaire Th\'eorique et Physique Math\'ematique, C.P. 229, Universit\'e Libre de Bruxelles (ULB), B-1050 Brussels, Belgium
}

\author{Christian Iliadis}
\email{iliadis@unc.edu}
\affiliation{University of North Carolina at Chapel Hill, Chapel Hill, NC 27599-3255, USA \\ and
Triangle Universities Nuclear Laboratory, Durham, North Carolina 27708-0308, USA}

\author{Richard Longland}
\email{richard\_longland@ncsu.edu}
\affiliation{North Carolina State University, Raleigh, NC 27695, USA\\ and
Triangle Universities Nuclear Laboratory, Durham, North Carolina 27708-0308, USA}

\begin{abstract}
Primordial or big bang nucleosynthesis (BBN)  is one of the three historical strong evidences for the big bang model. 
Standard BBN is now a parameter free theory, since the baryonic density of the Universe has been 
deduced with an unprecedented  precision from observations of the anisotropies of the cosmic microwave background (CMB) radiation.
There is a good agreement between the primordial abundances of \hli\ deduced from observations 
and from primordial nucleosynthesis calculations. 
However, the \sep\ calculated abundance is significantly higher  than the one 
deduced from  spectroscopic observations and remains an open problem.
In addition, recent deuterium observations have drastically reduced the uncertainty on D/H, to reach a value of 1.6\%. 
It needs to be matched by BBN predictions whose precision is now limited by thermonuclear reaction rate uncertainties.
This is especially important as many attempts to reconcile Li observations with models lead to an increased D prediction. 
Here, we re-evaluates the \dpg, \ddn\ and \ddp\ reaction rates that govern deuterium destruction, incorporating
new experimental data and carefully accounting for systematic uncertainties. Contrary to previous evaluations, we use 
theoretical ab initio models for the energy dependence of the \sfac{s}.  As a result, these rates increase 
{\em at BBN temperatures}, leading to a reduced value of D/H = (2.45$\pm0.10)\times10^{-5}$ (2$\sigma$), in agreement with observations.
\end{abstract}
 \date{\today}
 \maketitle


\section{Introduction}

The standard hot  big--bang model is supported by  three pieces of observational evidence: the cosmic expansion (the Hubble law), 
the Cosmic Microwave Background (CMB) 
radiation, and primordial or big bang nucleosynthesis (BBN).  
There is a good agreement between primordial abundances of \hli\ deduced from observations and from primordial
nucleosynthesis calculations. 
It is worth remembering that BBN has been essential in the past, first to estimate the baryonic density 
of the Universe, and give an upper limit on the  number of neutrino families.
The number of light neutrino families was later determined from the measurement of the $Z^0$ 
width by LEP experiments at CERN. The observations 
of the anisotropies of the cosmic microwave background by
WMAP \cite{WMAP9}, and more recently the Planck \cite{Planck13,Planck15} space missions, enabled the extraction of cosmological 
parameters  and, in particular, the baryonic density of the Universe. It was the last free parameter in BBN 
calculations, now measured with an uncertainty of less than 1\%:  \ob\   = 0.02225$\pm$0.00016 \footnote{
We note \ob\ $\equiv$ \obh, with $\Omega_{\mathrm{b}}$ the ratio of the baryonic to critical density and $h$ the  
Hubble constant in 100~km s$^{-1}$ Mpc$^{-1}$ units. 
} \cite{Planck15}. 
Higher precision standard BBN predictions are now needed for comparison with primordial abundances deduced from observations. 

Calculations of the \qua\ primordial abundance are in agreement with those deduced from observations in H{\sc ii} (ionized hydrogen) 
regions of compact blue galaxies \cite{Ave15}.
Contrary to \qua, \tro\ is both produced and destroyed in stars and thus
its abundance evolution as a function of time is not well known and difficult to compare with predictions.
It is well known that BBN calculations of \sep\ \cite{Cyb08b,CV10,Coc14b,Cyb15} overpredict the observations by a factor of $\approx$3. 
This is the so-called ``lithium problem", which has not found a satisfactory solution yet \cite{Fie11,Coc14a} (see also Ref.~\cite{Fu15}).
Promising ideas revolve around stellar physics or exotic physics, now that a nuclear physics solution is highly unlikely  \cite{Ham13}.
Deuterium's most primitive abundance is determined from the observation of very few
cosmological clouds at high redshift, on the line of sight of distant quasars. 
Recent observations of damped Lyman-$\alpha$ (DLA) systems at high redshift show a very small dispersion of deuterium abundance values, 
leading to a 1.6\% uncertainty on the mean value that is marginally compatible with BBN predictions.

Here, we 
will focus on the re-evaluation of the most important BBN reaction rates for deuterium nucleosynthesis.
Sensitivity studies (e.g., Ref.~\cite{Cyb04,CV10}) have shown that the \ddn, \ddp\ and \dpg\
reactions are the most influential for the D/H predicted abundance: a 10\% variation of their rates induces a relative variation of 
$-5.5$\%, $-4.6$\% and $-3.2$\%, respectively, of D/H. Concerning these reactions, since the most recent dedicated BBN evaluations of reaction 
rates \cite{Cyb04,Ser04,Des04}, new experiments were performed \cite{Leo06,Tum14}.  
On the contrary, no new experiment concerning the \dpg\
reaction has been performed and its rate uncertainty (5\%--8\% \cite{Des04,Ade11}), according to  Di Valentino \etal\ \cite{DiV14},
now dominates the error budget of D/H predictions. 

The recent NACRE--II 
 \footnote{In the following, we use ``NACRE'' when referring to the Angulo \etal\ \cite{NACRE} original evaluation,
and ``NACRE--II''  when referring to the recent sequel by Xu \etal\ \cite{NACRE2}.} 
\cite{NACRE2} evaluation provides new rates for these reactions. 
However, too few explanations are given regarding the data selection, fitting, and uncertainty estimation. 
Therefore, the published evaluated rates of these reactions are not suited to reach the precision
required for BBN calculations.   
Here, we re--evaluate the d+d rates, to take advantage of the new precise measurement by  Leonard \etal\ \cite{Leo06}, together with the \dpg\ rate.  
We use these new rates to derive BBN abundances and associated uncertainties.
We will then compare our new BBN predictions for deuterium with high-redshift observations in the framework of cosmic evolution models.

For our re-evaluations, we chose a compromise between adopting of the most recent, and more precise measurements 
only, (i.e., LUNA \cite{Cas02} for \dpg\ and Leonard \etal\ \cite{Leo06} for d+d) on the one hand, and including all available
experimental data in the fit on the other hand. The main difficulty in this analysis is the treatment and extraction of systematic uncertainties.  
Another difficulty is the choice of the fitting functions: polynomials 
\cite{NACRE,Cyb04}, splines \cite{Nol00} or R--Matrix \cite{Des04} have been used. 
For these three reactions we chose instead, as 
fitting functions, results from nuclear reaction models. It has the advantages of smoothing the accidental
fluctuations in experimental data and providing a better interpolation of the data.  In the case of a single data set, the fitting 
process is reduced to a normalization, but when several data sets have to be considered, a global normalization is required, which is discussed 
in the Appendices. In addition,  we found that, by using the theoretical ratio of \ddn\  and \ddp\ cross sections, it was possible to identify inconsistent 
data sets in an objective way.

This article is organized as follows. 
In Sections II and III, we discuss the normalization of \dpg, \ddn\ and \ddp\
theoretical \sfac{s} to experimental data. (The normalization method is presented in Appendix A and the experimental
data are discussed in Appendices B and C.)  In Section IV, we present the new reaction rates (tabulated in Appendix D) 
and the Monte Carlo
method for nucleosynthesis calculations. BBN results are discussed in Section V, while the cosmic deuterium evolution
is presented in Section VI. Finally, we show in Section VII that the new precise D/H observations put a strong constrain
on the proposed solution to the lithium problem. 

\section{The D\lowercase{(p},$\gamma)^3$H\lowercase{e} \sfac}
\label{s:dpg}

The sensitivity of the D/H abundance ratio to \dpg\ rate variations is \cite{CV10}
\begin{equation}
\frac{\Delta{\rm (D/H)}}{\rm D/H}=-0.32\frac{\Delta\langle\sigma{v}\rangle_{\mathrm{d(p,}\gamma)^3\mathrm{He}}}
{\langle\sigma{v}\rangle_{\mathrm{d(p,}\gamma)^3\mathrm{He}}}.
\end{equation}
Therefore, a precision of $\lesssim$ 5\% is required for the rate, to match the 1.6\% uncertainty in the observed value.
In the appendices, we detail our choice of the data sets we included in our analysis. Data sets for which no systematic uncertainty was quoted
(or when the quoted uncertainty was too large) were excluded from the fit. Nevertheless they are reported in the figures and tables,  
where the scatter of values gives an idea of their systematic uncertainties.

NACRE \cite{NACRE} 
used data from Refs.~\cite{Gri62,Gri63,War63,Ber64,Fet65,Ste65,Gel67,Wol67,Sch95,Ma97},
plus a  few high energy experiments and a polynomial fit, while Descouvemont \etal\ \cite{Des04} (DAACV hereafter) used a slightly different set of data, from 
Refs.~\cite{Cas02,Gri62,Gri63,War63,Wol67,Ma97,Bai70,Sko79,Sch97} and included
the data from Ref. \cite{Cas02} (post NACRE; from LUNA) with an R--matrix fit.
Figure~\ref{f:dpg} summarizes all the experimental data that we collected (see Appendix~\ref{a:dpg}), together with the fitted curves from Refs. \cite{Des04,Ade11}
and the theoretical \sfac\ from Marcucci \etal\ \cite{Marcucci}.
This is the theoretical \sfac\ that we renormalize to the data, as described in Appendix~\ref{s:norm}: the renormalization factor ($\alpha$) and the associated 
uncertainty ($\Delta\alpha$) are obtained by $\chi^2$ minimization [see Eqs.~(\ref{q:alpha}) and (\ref{q:dalpha})].
The results ($\alpha\pm\Delta\alpha$) of our analysis for the nine data sets \cite{Gri62,Gri63,War63,Wol67,Bai70,Ma97,Sch97,Cas02,Bys08b} can 
be found in Table~\ref{t:norm1}, column 3. Column 5 of the same table list the systematic uncertainties ($\epsilon$), only available for the four most recent 
data sets, to which we restrict our subsequent analysis. The systematic uncertainties are quadratically added to the \nfac\ uncertainties [Eq.~(\ref{q:epsi})]    
before calculating the recommended \nfac.

\begin{figure}[htb]
\begin{center}
 \includegraphics[width=.5\textwidth]{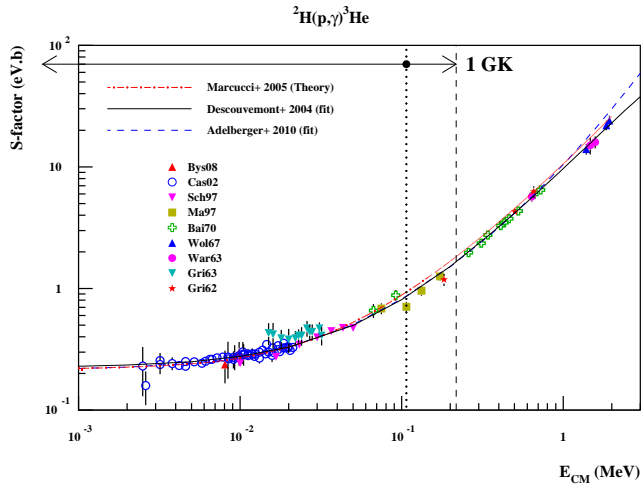} 
\caption{\sfac\ data considered in our evaluation compared to previous fits \cite{Des04,Ade11} or theory \cite{Marcucci}.
(See references in Table~\ref{t:norm1}.) The arrows, dashed and dotted vertical lines correspond to the Gamow window
at a temperature of 1~GK.}
\label{f:dpg}
\end{center}
\end{figure}

\begin{table}[htb]
\caption{\label{t:norm1} Results for the d(p,$\gamma$)$^3$He reaction.} 
\begin{center}
\begin{tabular}{c|c|c|c|c}
\hline\noalign{\smallskip}
\noalign{\smallskip}\hline\noalign{\smallskip}
 & & \multicolumn{3}{c}{\dpg}     \\
Ref. &  $N$ & $\alpha$ &  $\chi^2_\nu$ & $\epsilon$  \\ 
\noalign{\smallskip}\hline\noalign{\smallskip}
Bys08\cite{Bys08b}  & 3 &1.0365$\pm$0.1457 &  0.1360 & $\leq$0.08\\
Cas02\cite{Cas02} & 51 & 1.0243$\pm$0.0092 & 0.5792 & $\approx$0.045\\
Sch97\cite{Sch97}  & 7 & 0.9657$\pm$0.0062 & 11.1799 & 0.09  \\
Ma 97\cite{Ma97}  & 4 & 0.8469$\pm$0.0381 & 1.1052 & 0.09 \\
Bai70*\cite{Bai70}  & 11 & 0.9108$\pm$0.0143 & 0.3874 & n.a. \\
Wol67*\cite{Wol67}  & 3 & 0.9202$\pm$0.0514 & 0.2967 & n.a. \\
War63*\cite{War63}  & 3 & 0.8867$\pm$0.0581 & 0.2994& n.a. \\
Gri63*\cite{Gri63}  & 12 & 1.1749$\pm$0.0535 & 0.2322 & n.a. \\
Gri62*\cite{Gri62}  & 3 & 0.9104$\pm$0.0374 & 1.7730 & n.a. \\
\noalign{\smallskip}\hline\noalign{\smallskip}
\noalign{\smallskip}\hline
\end{tabular}\\
$\alpha$ = normalization factor, $N$ = number of data point and $\epsilon$ = systematic uncertainty.
Data sets marked with an asterisk have not been used in the analysis because the evaluation of systematics is not available.
\end{center}
\end{table}
\begin{figure}[h]
\begin{center}
 \includegraphics[width=.5\textwidth]{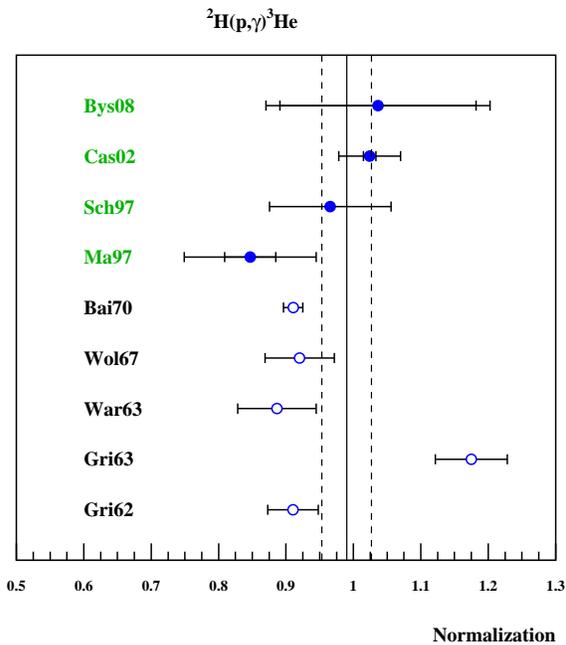} 
\caption{Normalization factors ($\alpha$) of the theoretical \dpg\ \sfac\ for different experiments obtained by 
Eq.~(\ref{q:alpha}) (blue circles). Full circles (green labels) correspond to data sets for which systematic uncertainties are available and are
selected in the evaluation. Other data sets (open circles) are shown for comparison only, 
and have not been used to derive our recommended average normalization factor. Their error bars correspond to  
uncertainties on the fit  [Eq.~(\ref{q:dalpha})] only. For the selected data sets (full circles) the error bars that include 
systematic uncertainties  [Eq.~(\ref{q:epsi})] are superimposed. 
Vertical lines correspond to the 
average value [Eq.~(\ref{q:amean})] and associated uncertainty [Eq.~(\ref{q:emean})]. 
(References can be found in Table~\ref{t:norm1}.) 
} 
\label{f:norm1}
\end{center}
\end{figure}

Figure~\ref{f:norm1} displays the \nfac{s} $\alpha$ from Table~\ref{t:norm1}, with error
bars that incorporate the systematic uncertainty [Eq.~(\ref{q:epsi})] when available. 
One observes that when the systematic uncertainties on normalization are included for the four selected data sets,
the dispersion of values of \nfac{s} becomes compatible with the error bars. 
Hence, a simple weighted average [Eq.~(\ref{q:amean})] and associated uncertainty [Eq.~(\ref{q:emean})] seems to us sufficient,
leading to $\alpha=0.9900\pm0.0368$ (Fig.~\ref{f:norm1}.) with a reduced chi--square of $\chi^2_\nu$= 0.71. 
We checked that the method used in some other evaluations (e.g., Ref.~\cite{Ser04}), and discussed in Appendix A.2,  
gives a very close value, $\alpha=0.9844\pm0.0366$  [from Eq.~(\ref{q:chi2o}) minimization].

\begin{figure}[htb]
\begin{center}
 \includegraphics[width=.5\textwidth]{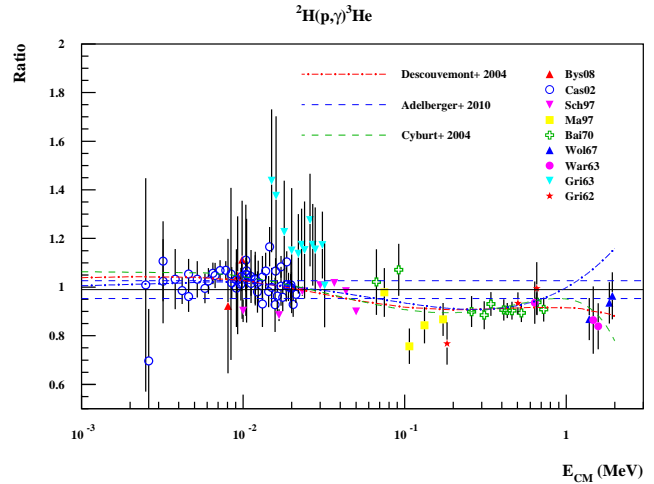} 
\caption{
Ratio of experimental and fitted \sfac{s} to the theoretical one \cite{Marcucci}. 
The horizontal lines correspond to the theoretical \sfac\ \cite{Marcucci} normalized to a subset of the experimental data
and the associated uncertainty (i.e. $\alpha\pm\Delta\alpha=0.9900\pm0.0368$).
Ratio of previous fits \cite{Des04,Cyb04} are driven below theory by the scarce data at BBN energies.
}
\label{f:dpgr}
\end{center}
\end{figure}

The zero energy theoretical \sfac\ is given by  $S(0)=0.21545$ eV$\cdot$b, which, after renormalization 
($\alpha=0.9900\pm0.0368$), leads to
$S(0)=0.213\pm0.008$ eV$\cdot$b, in excellent agreement with the value $S(0)=0.214^{+0.017}_{-0.016}$ eV$\cdot$b determined 
by Adelberger et al \cite{Ade11}. 
Hence, experimental data do not favor a {\em global} increase by a factor of  $\approx1.10\pm0.07$ as proposed by Di~Valentino \etal\ \cite{DiV14}
and the {\it Planck} collaboration \cite{Planck15} to better reproduce the Cooke \etal\ \cite{Coo14} deuterium observations (see \S~\ref{s:deut}).
Even when considering the experimental data that were not included in our fit (because their systematic uncertainties were not
quantified; see Fig.~\ref{f:norm1}), there is no evidence for such a global enhancement. 
Figure~\ref{f:dpgr} displays the experimental \sfac\ data, divided by the theoretical one \cite{Marcucci}. 
Except for Griffith \etal\ \cite{Gri63}, low energy data are in excellent agreement with our recommended average (horizontal solid lines). 
At higher energies, previous phenomenological fits \cite{Des04,Cyb04,Ade11} closely follow the experimental data points (Fig.~\ref{f:dpgr}).
In particular they are attracted by the Bailey et al. data \cite{Bai70} with very small error bars (Fig.~\ref{f:norm1}) but unknown systematic
uncertainty (see Appendix~B). For this reason we do not use these data in our fit. As it is based on a theoretical model \cite{Marcucci} that predict
the shape of the \sfac, our fit is little influenced by the Ma et al. data \cite{Ma97} with relatively large uncertainties (systematic uncertainties 
are not shown in Fig.~\ref{f:dpgr} but displayed in Fig.~\ref{f:norm1}). 
This explains that at BBN energies, the scarce data generally fall below our recommended average. 
Hence, while at low energy, our \sfac\ is in excellent agreement with other evaluations, {\em at BBN energies} our recommended \sfac\ is higher.
As a result, the BBN deuterium production calculated with our rate will be reduced (see \S~\ref{s:heli}). 
Precise cross section measurements at BBN energies ($\approx$100~keV) are hence strongly needed.

\section{The D\lowercase{(d,n)}$^3$H\lowercase{e} and D\lowercase{(d,p)}$^3$H \sfac{s}}
\label{s:dd}

The sensitivity of the D/H abundance ratio to \ddp\ and \ddn\ rate variations is \cite{CV10}
\begin{equation}
\frac{\Delta{\rm (D/H)}}{\rm D/H}=-0.54\frac{\Delta\langle\sigma{v}\rangle_{\mathrm{d(d,n)}^3\mathrm{He}}}{\langle\sigma{v}\rangle_{\mathrm{d(d,n)}^3\mathrm{He}}}
                                                   -0.46\frac{\Delta\langle\sigma{v}\rangle_{\mathrm{d(d,p)}^3\mathrm{H}}}{\langle\sigma{v}\rangle_{\mathrm{d(d,p)}^3\mathrm{H}}}
\end{equation}
so that a precision of better than 2\% is required for these rates.


Data from Refs.~\cite{Sch72,Gre95,Bro90,Kra87} were considered by the NACRE \cite{NACRE} collaboration and were also used in the 
R--matrix evaluation  of DAACV \cite{Des04}. 
Since DAACV, new measurements were performed by Leonard \etal\ \cite{Leo06} and by Tumino \etal\ \cite{Tum14}. 
Figures~\ref{f:ddn} and \ref{f:ddp} display all the experimental data that we collected 
(see Appendix~\ref{a:dd}). They show that the new, directly measured data \cite{Leo06} (labelled ``Leo06'' in Figures) follow reasonably well the DAACV R--matrix fit, 
even though it was calculated before the experiment was conducted.  
These figures also display the results from an {\it ab initio}
calculation by Arai \etal\ \cite{Ara11}, which we normalize to the experimental data as described in Appendix~\ref{s:norm}. 
This microscopic calculation uses a four-nucleon model with a realistic nucleon-nucleon
interaction. It was shown that the tensor force plays an important role in the d+d reactions.
However, the theoretical work of Ref.~\cite{Ara11} was focused on low energies, and only partial waves up to $J=2$
have been included. For this reason, above $~1$ MeV, the theory underestimates the data. 
Consequencely, we choose to limit the normalization to data below 0.6~MeV, which is well above the energy region important for BBN (dashed vertical lines). 

\begin{figure}[htb]
\begin{center}
 \includegraphics[width=.5\textwidth]{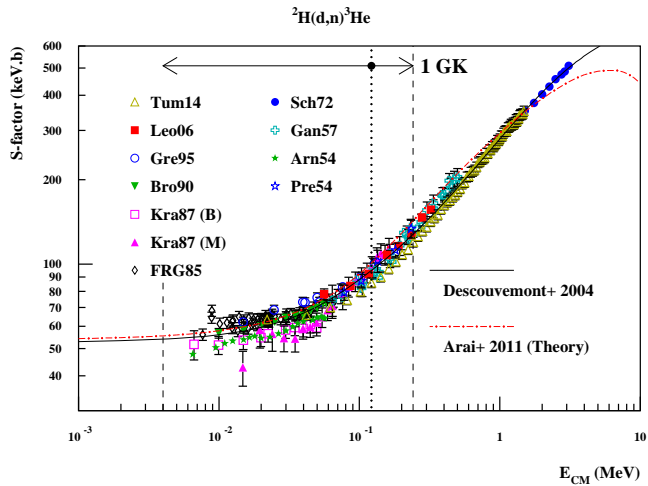} 
\caption{Experimental data considered for \ddn\ rate calculations, compared to DAACV R--Matrix  fit and theory \cite{Ara11}.
The horizontal arrows indicate the Gamow window at 1~GK.
(See references in Table~\ref{t:norm2}).}
\label{f:ddn}
\end{center}
\end{figure}

\begin{figure}[htb]
\begin{center}
 \includegraphics[width=.5\textwidth]{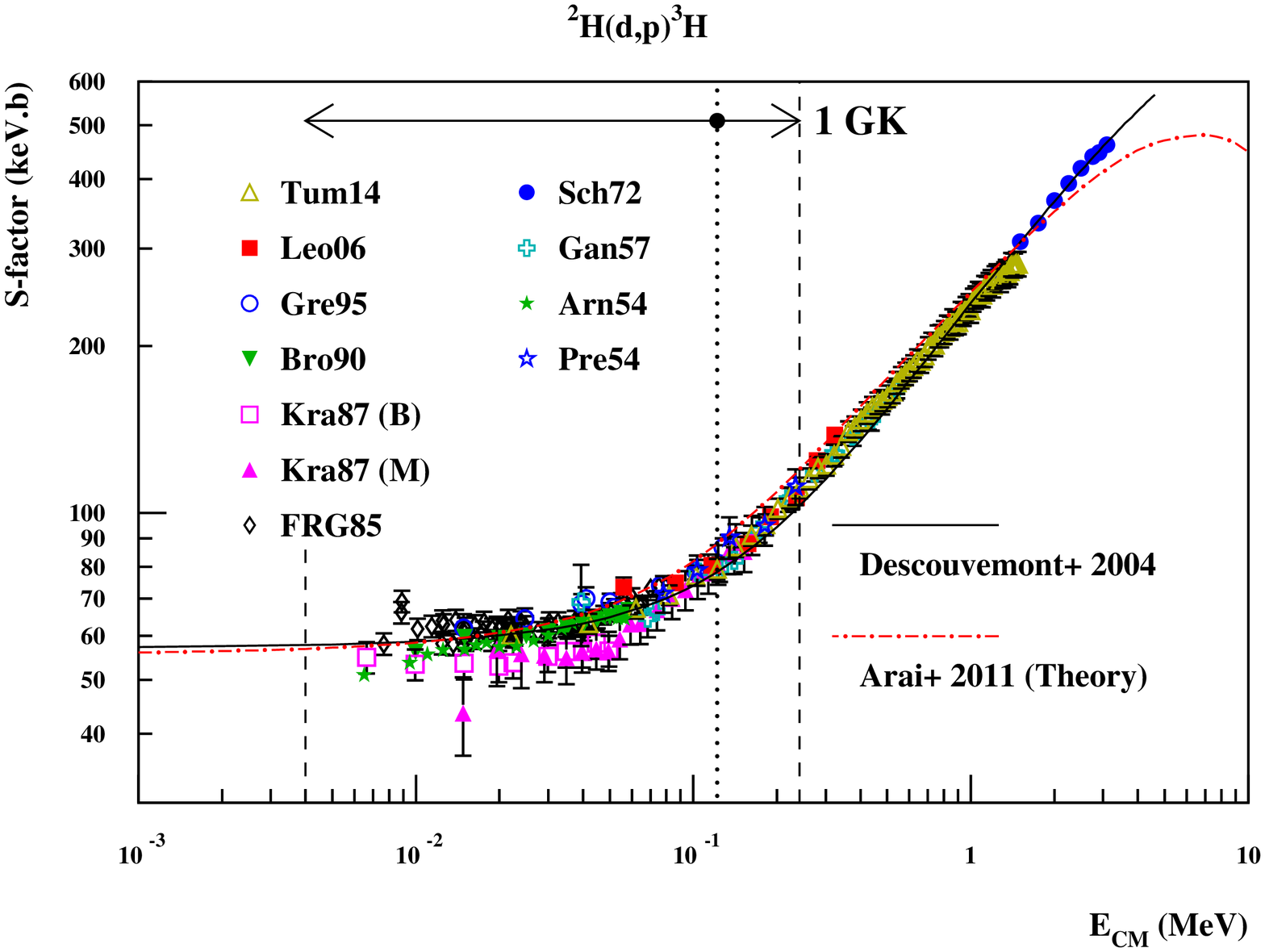} 
\caption{Same as Fig.~\ref{f:ddn}, but for \ddp.}
\label{f:ddp}
\end{center}
\end{figure}

\begin{figure}[htb]
\begin{center}
 \includegraphics[width=.5\textwidth]{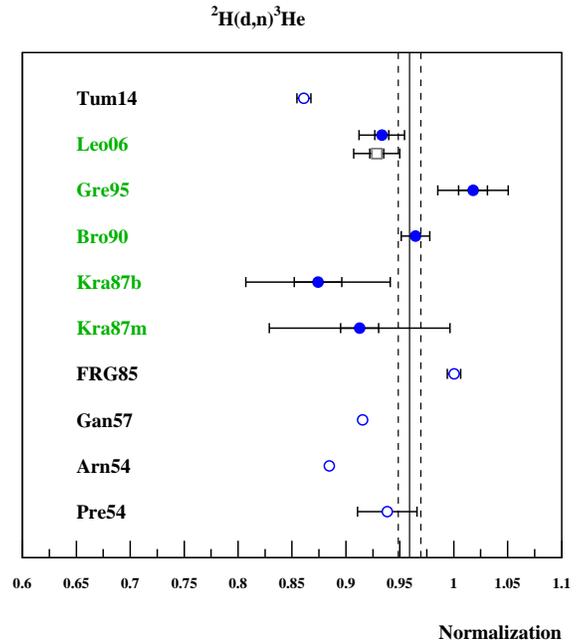} 
\caption{Same as Fig.~\ref{f:norm1}, but for the \ddn\ reaction: only data shown by blue solid circles and green labels are used
for the final normalization. Those selected data also display double error bars: the uncertainties from the fit and the total uncertainty including 
systematic uncertainties.}
\label{f:norm2n}
\end{center}
\end{figure}

\begin{figure}[htb]
\begin{center}
 \includegraphics[width=.5\textwidth]{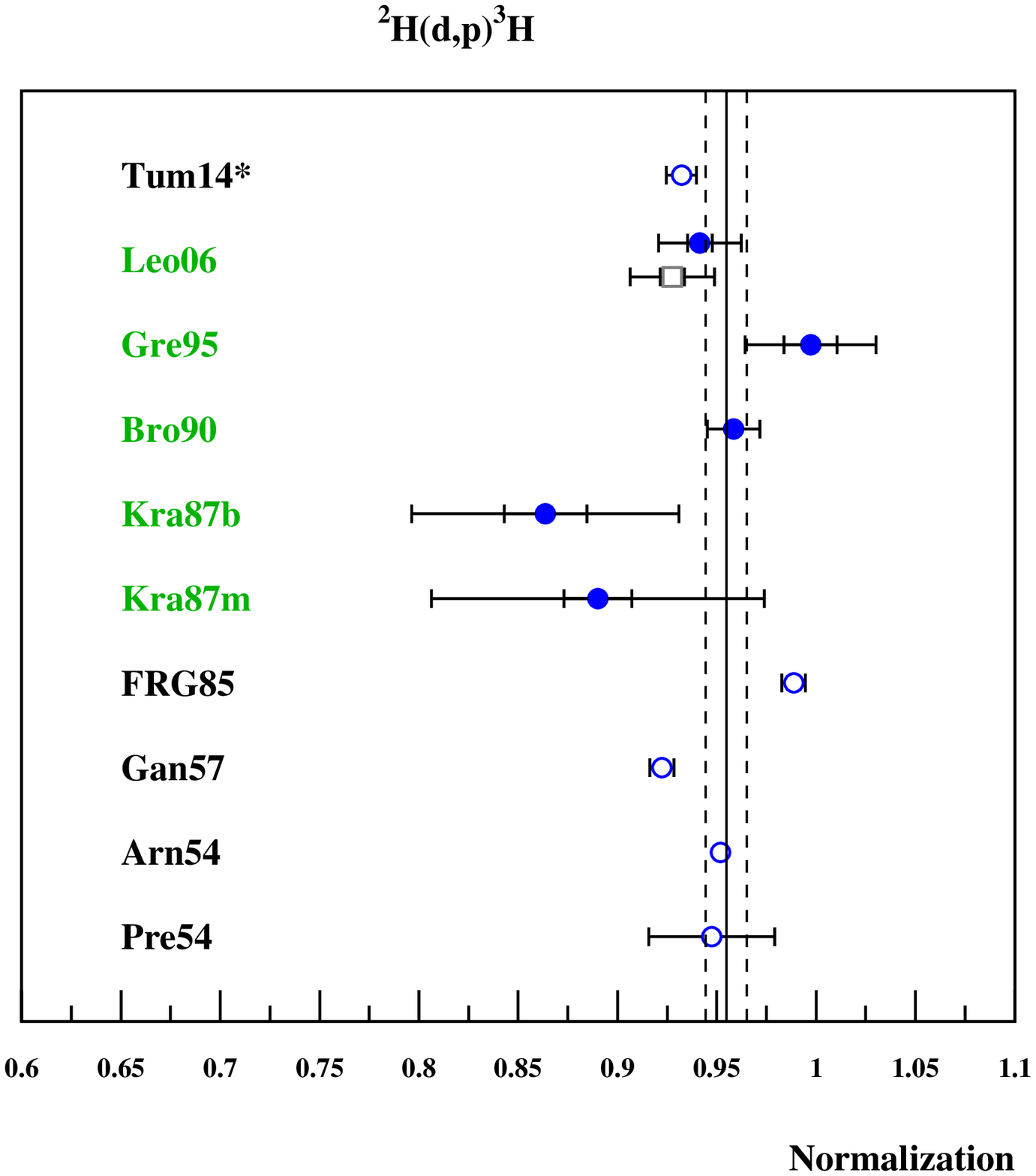} 
\caption{Same as Fig.~\ref{f:norm1}, but for the \ddp\ reaction.}
\label{f:norm2p}
\end{center}
\end{figure}

Similar to our analysis of the \dpg\ reaction (Section \ref{s:dpg}) we assume different normalization factors in different experiments and allow them to be different for
\ddn\ and \ddp.
The results of our fits for the collected eleven data sets 
\cite{Pre54,Arn54,Gan57,Sch72,first,Kra87,Bro90,Gre95,Leo06,Tum14}  (Appendix~\ref{a:dd}) can be found in Table~\ref{t:norm2}
and in Figures~\ref{f:norm2n} and \ref{f:norm2p}. 
Because of the limited energy range, 0.015~MeV $\leq{E}\leq 0.6$~MeV (considering electron screening at low energy and nuclear model restrictions at high energy,
see Appendix~\ref{a:dd}), the number of adopted data points, $N$, is smaller compared to the original publications. 
For instance, because of these limitations, we had to disregard the 
Shulte \etal\ \cite{Sch72} data and the lowest energy Krauss \etal\ \cite{Kra87} data. 

In all experiment but one, we fitted the \ddn\ and \ddp\ data sets independently. 
Nevertheless, with a few exceptions, in a given experiment, the \ddn\ (Fig.~\ref{f:norm2n}) and \ddp\ (Fig.~\ref{f:norm2p})  \nfac{s}  
are very similar.
For the Leonard \etal\ data \cite{Leo06}, we took advantage of the published error matrix and performed a fit
taking into account all correlations between data points of different energies or reactions. 
This resulted in a simultaneous fit of both the \ddn\ and \ddp\ cross sections.
The results, displayed in 
Fig.~\ref{f:norm2n} and Fig.~\ref{f:norm2p} as grey squares show little difference with the simple fit (blue) circles 
and are not used.  
As for the \dpg\ reaction, we select the data sets for which the systematic uncertainties are published (see last column
of Table~\ref{t:norm2}).  As discussed in more detail in Appendix C, we do not use the indirect measurement from
Tumino et al. \cite{Tum14} as the energy dependence of their experimental \sfac{s} is slightly different from
theory and other experiments, in particular for the \ddn\ reaction as it can be seen in Figs.~\ref{f:ddn} and \ref{f:norm2n}. 

\begin{figure}[htb]
\begin{center}
 \includegraphics[width=.5\textwidth]{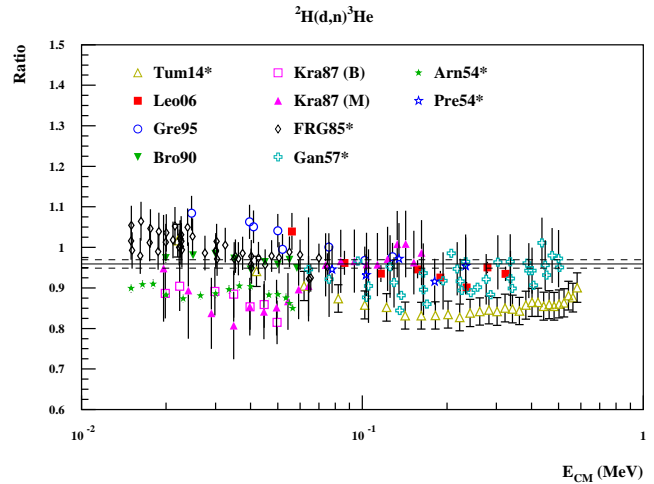} 
\caption{Same as Fig.~\ref{f:dpgr}, but for the \ddn\ reaction and 0.015~MeV $\leq{E}\leq$0.6~MeV.}
\label{f:rddn}
\end{center}
\end{figure}

\begin{figure}[htb]
\begin{center}
 \includegraphics[width=.5\textwidth]{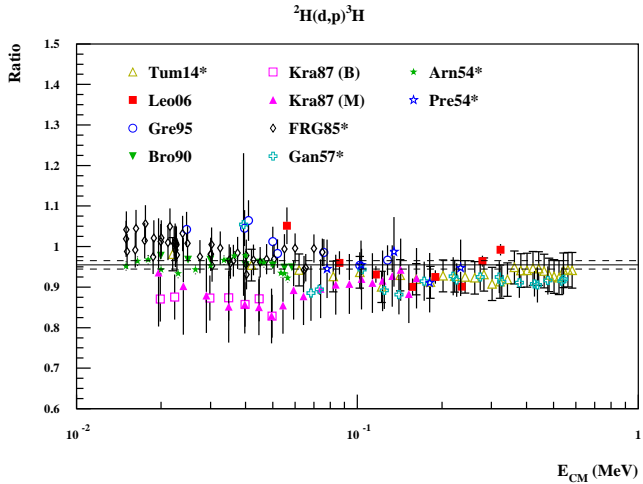} 
\caption{Same as Fig.~\ref{f:dpgr}, but for the \ddp\ reaction and 0.015~MeV $\leq{E}\leq$0.6~MeV.}
\label{f:rddp}
\end{center}
\end{figure}

For the recommended normalization factor, we obtain $\alpha=0.9590\pm0.0104$ and
$\alpha=0.9549\pm0.0103$ (weighted average error) for the \ddn\ and \ddp\  reactions with reduced chi--squares close 
to unity (1.33 and 0.92). (We obtained similar results, $\alpha=0.9579\pm0.0100$, and  $0.9541\pm0.0099$ when using the
alternative method presented in Appexdix A.2.) 
Figures~\ref{f:rddn} and \ref{f:rddp} display the scatter of all experimental data, in the 0.015~MeV $\leq{E}\leq$0.6~MeV range,
relative to the theoretical model from Ref. \cite{Ara11}. Data sets that significantly deviate from the fits were just not included
in the fit (e.g., Tumino et al. \cite{Tum14} and Arnold et al. \cite{Arn54} for reasons discussed in Appendix~C) or have large systematic uncertainties 
(Krauss et al. \cite{Kra87}) that are not included in the error bars of these figures.

\begin{table*}
\caption{\label{t:norm2} Results of normalization of individual data sets.} 
\begin{center}
\begin{tabular}{l|c|c|c|c|c|c}
\hline\noalign{\smallskip}
\noalign{\smallskip}\hline\noalign{\smallskip}
 & & \multicolumn{2}{c|}{\ddn} & \multicolumn{2}{c|}{\ddp} &       \\
Ref.  & $N$ & $\alpha$ &  $\chi^2_\nu$  & $\alpha$ & $\chi^2_\nu$   &  $\epsilon$ \\ 
\noalign{\smallskip}\hline\noalign{\smallskip}
Tum14* \cite{Tum14} & 29 & 0.8610 $\pm$ 0.0064 &1.0332 & 0.9322  $\pm$  0.0075  & 0.1582 & n.a. \\
Leo06 \cite{Leo06}& 8 &0.9333   $\pm$ 0.0065 & 2.0327 & 0.9415  $\pm$  0.0061 & 5.3758 & 0.02 \\
Gre95 \cite{Gre95} &8 &1.0158   $\pm$ 0.0134 &1.2472 &  0.9972 $\pm$   0.0134&  0.9989 &  0.03\\  
Bro90 \cite{Bro90} & 9&0.9644   $\pm$ 0.0025 & 2.3659 &0.9584  $\pm$  0.0020  & 1.9690 &  0.013\\   
Kra87 (B) \cite{Kra87} & 7&0.8683   $\pm$ 0.0220 & 0.2919&  0.8637  $\pm$  0.0208&  0.1001 & 0.064 \\   
Kra87 (M)  \cite{Kra87} & 20&0.9185  $\pm$  0.0177& 0.6236&  0.8902  $\pm$  0.0171&  0.1766&  0.082\\   
FRG85* \cite{first} & 45 &0.9913   $\pm$ 0.0062& 0.6509 & 0.9887  0.0060 &  0.5044&  n.a. \\   
Gan57* \cite{Gan57} & 36/18&0.9155  $\pm$  0.0013 & 22.8927 & 0.9223  $\pm$  0.0061&  0.7350&   n.a. \\   
Arn54* \cite{Arn54} & 15&0.8860  $\pm$  0.0027 & 2.6995&   0.9519  $\pm$  0.0029& 2.1919 & n.a. \\   
Pre54* \cite{Pre54} & 5&0.9384  $\pm$  0.0275& 0.0988&   0.9475  $\pm$  0.0317 & 0.1186 & n.a.  \\ 
\noalign{\smallskip}\hline\noalign{\smallskip}
\noalign{\smallskip}\hline
\end{tabular}
\end{center}
\end{table*}

\section{Reaction rates and uncertainties}

\begin{figure}[htb]
\begin{center}
\vskip -2.cm
 \includegraphics[width=.5\textwidth]{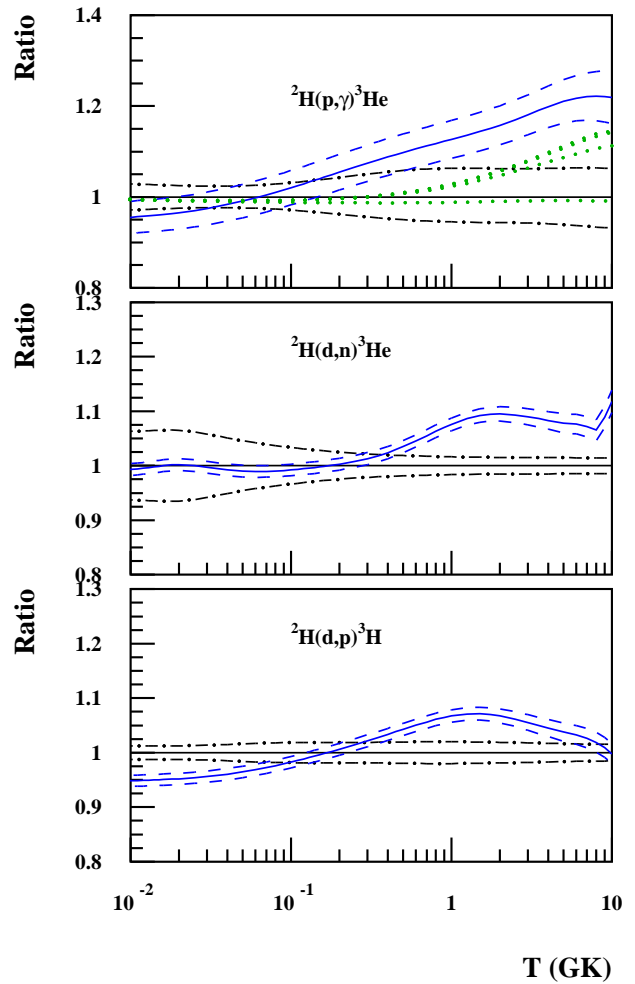} 
\caption{New rates (blue), compared with previous \cite{Des04} results (black), normalized to the DAACV \cite{Des04} 
recommended rate. In the top panel, green curves: our re-calculation of the \dpg\ reaction rate using the DAACV \sfac,
but with increased numerical integration limits.
}
\label{f:rates}
\end{center}
\end{figure}

The reaction rates were calculated by numerical integration of the theoretical \sfac{s}, after normalization.  
Above the energies imposed by the limitations of the models (2~MeV and 0.6~MeV,
respectively, for \dpg\ and d+d), the  \sfac{s} are supplemented by the DAACV \cite{Des04} results.  
The influence of this 
high energy \sfac\ merging is negligible at BBN temperatures, but allows for calculating the tabulated rates on a conventional
temperature grid. Figure~\ref{f:rates} shows the new rates compared to the DAACV  \cite{Des04} that were
used in previous works (e.g., Ref.~\cite{Coc14b}). 
In the upper panel (\dpg\ rate), we also display (dotted lines) the result of our numerical integration of the DAACV \sfac\ within 
different energy intervals $E_0\pm n \Delta E_0$, with  $n$ = 2, 3, 4 and 5,
where $E_0$ and $\Delta E_0$ define the Gamow window (e.g., Eqs. (5) and (6)   in NACRE \cite{NACRE}).
In DAACV the rate was calculated this $n$=2, not sufficient to reach the high precision needed here but that $n\gtrsim3$ is required.
Hence, to derive a more precise rate from our recommended \sfac{s}, we used a wider interval ($n$ = 4) in our calculations.
For each reaction, the 1$\sigma$ uncertainties on rates, $\mathrm{N_A}\langle\sigma{v}\rangle_{^\mathrm{high}_\mathrm{low}}$,
are obtained by using  the  1$\sigma$ uncertainties on the \nfac{s} ($\alpha\pm\Delta\alpha$)  
to rescale the theoretical \sfac{s}, or by using the DAACV uncertainty on the \sfac\ at the highest energies.
It is worth noting that these rate uncertainties are statistically defined (1$\sigma$ limits), at variance with the limits provided in
some other evaluations, e.g., NACRE \cite{NACRE} and NACRE--II  \cite{NACRE2}.
The recommended reaction rates $\mathrm{N_A}\langle\sigma{v}\rangle_\mathrm{rec}$ calculated from the \sfac{s}
rescaled by $\alpha$ 
can be found in Table~\ref{t:rates}, together with the uncertainty factors, ($f.u.$),
defined \cite{Eval1} as 
\begin{equation}
f.u.\equiv\sqrt{\mathrm{N_A}\langle\sigma{v}\rangle_{\mathrm{high}}/\mathrm{N_A}\langle\sigma{v}\rangle_{\mathrm{low}}}.
\label{q:fu}
\end{equation}
Except at the highest energies where DAACV rate uncertainties are used, one simply has
\begin{equation}
f.u. = \sqrt{(\alpha+\Delta\alpha)/(\alpha-\Delta\alpha)}\approx1+\Delta\alpha/\alpha.
\end{equation}

In the Monte Carlo calculations,  the reaction rates ${N_A}\langle\sigma{v}\rangle^k$, (with $k$ being the index of the reaction), 
are assumed to follow a lognormal distribution:
\begin{equation}
{N_A}\langle\sigma{v}\rangle^k=\exp\left(\mu_k(T)+p_k\sigma_k(T)\right)
\label{q:ln}
\end{equation}
where  $p_k$ is sampled according to a {\em normal} distribution of mean 0 and variance 1 (Eq.~(4)  of Ref.~\cite{Stats14}).
The $\mu_k$ and $\sigma_k$ determine the location of the distribution and its width. For the \dpg, \ddn\ and \ddp\
reactions, they are derived from the values listed in Table~\ref{t:rates}, using $\exp(\mu_k)\equiv\mathrm{N_A}\langle\sigma{v}\rangle^k_\mathrm{rec}$
and $\exp(\sigma_k)\equiv\left(f.u.\right)^k$.
From the resulting histograms of 
calculated abundances, the median and 68\% confidence interval is obtained from the 0.5, 0.16 and 0.84
quantiles \cite{Eval1,Stats14}.

\section{BBN results}
\label{s:heli}

The standard analysis of the CMB data by the Planck satellite showed that the temperature and polarization angular power 
spectra are consistent with a spatially- Euclidean $\Lambda$CDM cosmological model with a power-law initial power spectrum 
for adiabatic scalar perturbations. 
The analysis includes parameters describing the baryonic and CDM densities, the cosmological constant, the amplitude and spectral index 
of the primordial power spectrum and the optical depth due to reionization. Besides, the present photon temperature is well-measured 
to be $T_0=2.7255\pm0.0006$~K \cite{Fix09}. Assuming thermal equilibrium prior to neutrino decoupling, 
the radiation density is inferred by assuming that the effective number of neutrino families is $N_{\rm eff}=3.046$ \cite{Man05}. 
Among the various combinations of the temperature, $E$-polarization data and lensing of the CMB by the large scale structure of the universe 
(see Table~4 of Ref.~\cite{Planck15}), we adopt the constraints obtained with the largest set of data (TT,TE,EE+lowP) without any external data and 
without taking the lensing data into account. This gives a constraint on the baryonic density parameter \ob\ = 0.02225$\pm$0.00016 
with a 68\% confidence level
 \footnote{This corresponds to a number of baryons per photon of $\eta$ = (6.0914$\pm$0.04380)$\times10^{-10}$,
using $\eta=2.7377\times10^{-8}\times$~\ob\ \cite{Coc14b}.}.
In full generality, when combining data one should consider a consistent code predicting both the BBN abundances and the CMB 
angular power spectra. Note, in particular, that the latter requires us to determine the helium abundance which affects the recombination 
process since helium recombines before hydrogen. The Planck results \cite{Planck15} used a posterior mean of $Y_p\sim0.2453$ 
``predicted by BBN, with theoretical uncertainties dominating over the Planck error $\Omega_{\rm b}h^2$...." (see Table 4 of Ref.~\cite{Planck15}). 
One could introduce the new parameter $Y_p$ but it is not free since it is related to \ob\ through BBN. 
The Planck analysis uses the PArthENoPE code \cite{Pis08} assuming a neutron mean lifetime of 880.3~s. Indeed, 
what is meant by ``theoretical uncertainties" includes ``uncertainties in the neutron lifetime and a few nuclear reaction rates". 
This emphasizes the importance of the present analysis that uses, compared to thePArthENoPE code,  improved 
thermonuclear reaction rates  relevant for D (and Li) nucleosynthesis. 
As a first analysis, and in order to confront Planck results with our BBN predictions, we compare the independent predictions 
of the CMB and BBN with 68\% and 95\% confidence level. Such a preliminary approach is sufficient to identify if there is any 
tension between the two methods and to determine whether they are compatible at a given confidence level.

Table~\ref{t:heli} shows the step by step progression of the \hli\ abundances with improved input data (reaction rates and \ob). 
The first column (a) lists to the results of the Monte Carlo calculation from Ref.~\cite{Coc14b} 
(with \ob=0.02218$\pm$0.00026 and for $\tau_{\rm{n}}$ = 880.1$\pm$1.1~s) 
for comparison. 
The second column (b) uses the same 425--reaction network but for  \ob=0.02225 (and  $\tau_{\rm{n}}$ = 880.3~s).
The largest difference, on \sep, between the median of the Monte Carlo distribution of abundances (a), and the calculation using 
nominal values of parameters (b), is due to the non Gaussian shape of the Li/H abundance distribution, statistical fluctuations,
and to minute updates of physical constants. 
Two rates affecting $^7$Be nucleosynthesis have been updated since DAACV \cite{Des04}.
The $^3$He($\alpha,\gamma)^7$Be rate from  DAACV had been superseded by Cyburt \& Davids~\cite{Cyb08a}, 
who included new results from LUNA. An improved evaluation of the $^3$He($\alpha,\gamma)^7$Be reaction rate and 
associated uncertainty has been published \cite{deB14} since, using a Monte--Carlo based R--matrix analysis, and
can be considered up to date. 
Even though the $^7$Be(n,$\alpha)^4$He reaction cannot help solve the lithium problem, its rate was uncertain
and affected the \sep\ production at the few percent level.
Until recently, the only published rate came from an evaluation by Wagoner \cite{Wag69}. 
Very recently, Hou \etal\ \cite{Hou15} clarified the origin of the Wagoner rate, but more importantly have
re--evaluated it, based on $^4$He($\alpha$,n)$^7$Be, $^4$He($\alpha$,p)$^7$Li and 
$^7$Li(p,$\alpha)^4$He experimental data, using  charge symmetry and/or detailed balance principles. 
The next two columns (c and d) in Table~\ref{t:heli} show the effect of updating the $^7$Be(n,$\alpha$)\qua\ \cite{Hou15}
and  $^3$He($\alpha,\gamma)^7$Be  \cite{deB14} rates, respectively. 
The effects of using our new \ddn\  and \ddp\ and \dpg\  rates instead of DAACV \cite{Des04} are 
displayed in the next columns, (e) and (f) respectively. As expected from Fig.~\ref{f:rates}, the D/H abundance is significantly reduced,
together with a concomitant \sep\ increase.  
The results of a Monte Carlo calculation, performed as in Ref.~\cite{Coc14b}, but with the
updated rates and \ob\ = 0.02225$\pm$0.00016 \cite{Planck15} (see above)  and  $\tau_{\rm{n}}$ = 880.3$\pm$1.1~s \cite{PDG}
are shown in column (g), compared to observations (h).
For deuterium, we obtain
\begin{equation}
\mathrm{D/H} = (2.45\pm0.10)\times10^{-5}\;\;(2\sigma). 
\label{q:dh}
\end{equation}

\begin{table*}[htbp!] 
\caption{\label{t:hlix} Primordial abundances}
\begin{center}
\begin{tabular}{ccccccccc}
\hline
\hline
& a & b & c & d & e & f & {\bf g (predicted)} & h (observed)\\
\ob\  &  0.02218$\pm$0.00026  &  0.02225&  0.02225&  0.02225&  0.02225&  0.02225 &  {\bf 0.02225$\pm$0.00016} & \\
\hline
$Y_p$      & 0.2482$\pm$0.0003  &0.2482&0.2482&0.2482&0.2484&0.2484 & {\bf 0.2484$\pm$0.0002} & 0.2449$\pm$0.0040\cite{Ave15}\\
\deu/H   ($ \times10^{-5})$ &$2.64^{+0.08}_{-0.07}$&2.635&2.635&2.635&2.526&2.452& {\bf 2.45$\pm$0.05} & 2.53$\pm$0.04 \cite{Coo14}\\
\tro/H    ($ \times10^{-5}$)  & 1.05$\pm$0.03 &1.047&1.047&1.047&1.038&1.070& {\bf 1.07$\pm$0.03} & 1.1$\pm$0.2 \cite{Ban02}\\
\sep/H ($\times10^{-10}$)   &  $4.94_{-0.38}^{+0.40}$&5.040&5.102&5.131&5.343&5.651& {\bf 5.61$\pm$0.26} &   1.58$^{+0.35}_{-0.28}$\cite{Sbo10}\\
\hline
\hline
\end{tabular}\\
Ref.  \cite{Coc14b} (a) ; Update of \ob\ (b), $^7$Be(n,$\alpha$)\qua\ (c),
\hag\ (d), \ddn\ and \ddp\ (e), and \dpg\ (f) new rates, Monte Carlo (1$\sigma$) {\bf (g)} and observations (h).
\end{center}
\label{t:heli}
\end{table*}

\section{Cosmic deuterium evolution}
\label{s:deut}

Starting from our new BBN prediction [Eq.~\ref{q:dh}], 
it is interesting to follow the cosmic deuterium evolution. This isotope is a good tracer of stellar formation since it can only be destroyed after the BBN stage.

\subsection{Observations}

The primitive abundance of deuterium is determined from the observation of cosmological clouds at high redshift
located on the line of sight of distant quasars. Very few observations  are available so far.
Pettini and Cooke \cite{Pet12} and, more recently Cooke \etal\ \cite{Coo14} observed, or 
reanalyzed, five DLA systems at redshift 2--3 and  
derived a mean value D/H = (2.53$\pm$0.04)$\times10^{-5}$. 
Recently, Riemer-S{\o}rensen \etal\  \cite{Rie15} re--measured the $z$ = 3.256 absorption system towards the quasar 
PKS~1937$-$101 and have determined a robust value of D/H = (2.45$\pm$0.28)$\times10^{-5}$.
Finally Noterdaeme \etal\ \cite{Not12} measured D/H = (2.59$\pm$0.15)$\times10^{-5}$ at $z$ = 2.621 towards CTQ~247.
Our present BBN D/H calculated value of (2.45$\pm$0.10)$\times10^{-5}$  (2$\sigma$) is in agreement with these observational 
constraints, although the observations tend to be slightly higher.

The  D/H ratio can also be derived from observations of HD and H$_2$ molecules in DLAs assuming that chemistry does not affect its value.
The observed ratios take very different values, which may cast some doubt on this latter assumption \cite{Tum10}.
Srianand \etal\ \cite{Sri10},  Ivanchik \etal\ \cite{Iva10}, and Balashev \etal\ \cite{Bal10}  
measured D/H = (1.17$^{+ 0.49}_{-0.34})\times10^{-5}$,   (3.6$^{+ 1.9}_{-1.1})\times10^{-5}$ and (3.6$^{+ 1.9}_{-1.1})\times10^{-5}$ 
towards   J1337+3152,  Q~1232+082,  and both J~0812+3208 and Q~1331+170
 at  $z$~=~3.102,  2.3377, and 2.626 and 1.777, respectively.
Indeed,  Le Petit \etal\ \cite{LeP09} modeled the deuterium chemistry
and showed that the derived D/H ratio strongly depends on the initial physical conditions such as temperature and density. 
However, they considered dense clouds, whereas most DLAs are diffuse structures. 
Since the situation is not clear, it is premature to use these observational measurements to compare with the results of our models.

Recent local D/H observations added new constraints on the cosmic deuterium astration factor, $f_D$, which is 
defined as the ratio of the BBN to the present deuterium abundances, D$_\mathrm{BBN}$/D$_\mathrm{present}$.
In the local  interstellar medium (ISM), Prodanovi\'c \etal\ \cite{Pro10a} find their best estimate for the undepleted ISM deuterium abundance 
to be D/ H = (2.0$\pm$0.1)$\times10^{-5}$, leading to  $f_D<1.26\pm$0.1. 
In the  local galactic disk, Linsky \etal\ \cite{Lin06} analyzed spectra obtained with the Far Ultraviolet Spectroscopic Explorer  (FUSE ) satellite, 
together with spectra from the Copernicus and interstellar medium absorption profile spectrograph (IMAPS) instruments. 
This study reveals a very wide range in the observed D/H ratio. Spatial variations in the depletion of deuterium in dust grains could
explain these local variations. Finally, they argue that the most representative value for the D/H ratio within 1~kpc of the Sun is
(2.31$\pm$0.24)$\times10^{-5}$. 
The deuterium astration factor, $f_D$, is in this context less than 1.1.  
Finally, Savage \etal\ \cite{Sav07} use high-resolution ultraviolet spectra in the lower galactic halo and obtain 
D/H = $(2.2^{+ 0.8}_{-0.6})\times10^{-5}$. This value is consistent with the results mentioned above, but with a very large error bar.

\subsection{Evolution}

\begin{figure}[htb]
\begin{center}
 \includegraphics[width=.5\textwidth]{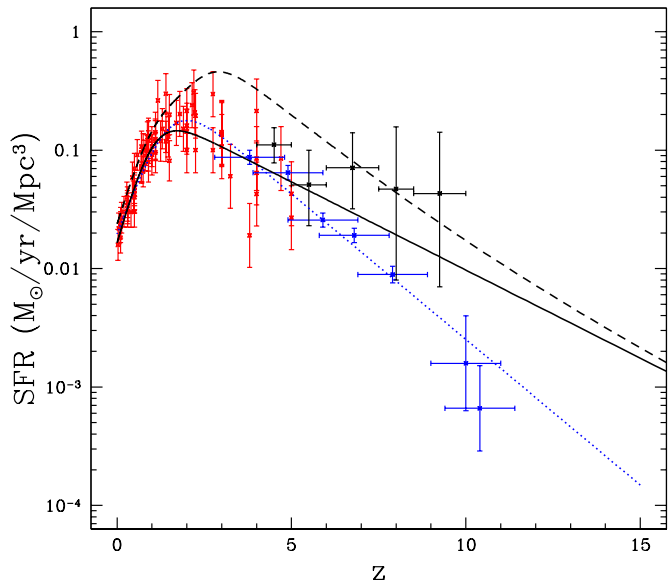} 
\caption{Cosmic SFR as a function of redshift. The solid black line fit from Trenti \etal\ \cite{Tre13}  and 
Behroozi and Silk \cite{Beh14}  (see also Vangioni \etal\ \cite{Van14}) is adopted in the present work. 
The dashed black curve corresponds to an upper limit of the SFR, an extreme case obtained by 
adding an intermediate mass SFR component (between 2 and 8 $M_\odot$) to maximize the deuterium destruction.
For comparison, the dotted blue line fits the
observations coming from high redshift galaxy surveys \cite{Beh13} (red points) and \cite{Bou14,Oes14} (and references therein) (blue points). 
The exact slope of the SFR at high redshift has little impact on the deuterium evolution.}
\label{f:sfr}
\end{center}
\end{figure}

We now consider the cosmic evolution of D/H in a cosmological context  in the light of the new, somewhat low, deuteriumm primordial value
derived here. 
It is well known that, due to its fragility, deuterium is destroyed during the cosmic evolution  (as soon as $T >10^{5}$~ K). 
In this context, we follow the cosmic chemical evolution using a model developed in Refs.~\cite{Dai06,Rol09,Van14}, based 
on a hierarchical model for structure formation \cite{Pre74,Wyi03}. 
A key ingredient to all evolution models is the global cosmic star formation rate, SFRs (a specific analysis devoted to different SFR is 
performed in Ref.~\cite{Van14}), whose evolution with redshift is constrained by many observations. 
Recent data from high redshift galaxy observations  (the Hubble Ultra Deep Field) have significantly extended the range of redshifts for its 
determination, from $z$ = 4 up to 10 \cite{Bou14,Oes14}. Figure~\ref{f:sfr} shows the SFR fit using these observations (blue points and  blue dotted line).
On the other hand, observations of high $z$ gamma--ray bursts (GRBs) tend to favor a large amount of yet unobserved SFR at 
$z>$ 9 \cite{Kis13} (black points and black solid line). 
Extracting the SFR from the GRB rate is not free from uncertainties and biases \cite{Tre13}. In the following, we will use
the SFR derived from Ref.~\cite{Tre13} (see Figure~\ref{f:sfr}, black line), which is consistent with other observational 
constraints as shown by Vangioni \etal\ \cite{Van14}.  
To estimate the maximum astration factor of D, we have also considered an extreme 
case, adding an intermediate mass SFR component (between 2 and 8 $M_\odot$) which is shown in  Fig.~\ref{f:sfr} by the black dashed curve.
Note that the exact slope of the SFR at high redshift has no impact on the deuterium evolution (contrary to heavier elements). 
Indeed, it is well known that deuterium  destruction is governed by low mass stars (since the gas is essentially trapped in these stars), 
whereas metallicity production (elements other than H and He) is governed by high mass stars, which, having  short lifetimes, 
start rejection of enriched matter at high redshift.
A weak destruction of deuterium is consequently not incompatible with a significant formation of heavy elements. 

Assuming a given cosmic evolution of the star formation rate (SFR), the model follows the evolution of the
baryons  abundance in stars, in diffuse structures (interstellar medium, ISM) and in the intergalactic medium (IGM).
The model includes a description of mass exchanges between the IGM and ISM (structure formation, galactic outflows), and between the ISM 
and the stellar component (star formation, stellar winds and supernova explosions). Once the cosmic SFR is specified, several quantities 
are obtained as a function of the redshift, namely the abundances of chemical elements, and more specifically deuterium.
We consider for the present study the results of the best model described in Ref.~\cite{Van14}, including a standard mode of Population II/I star formation between 
0.1~$M_\odot$ and 100~$M_\odot$. The initial mass function (IMF) slope is set to the Salpeter value, i.e., $x=1.35$ \cite{Sal55,Cha14}. 

\begin{figure}[htb]
\begin{center}
 \includegraphics[width=.5\textwidth]{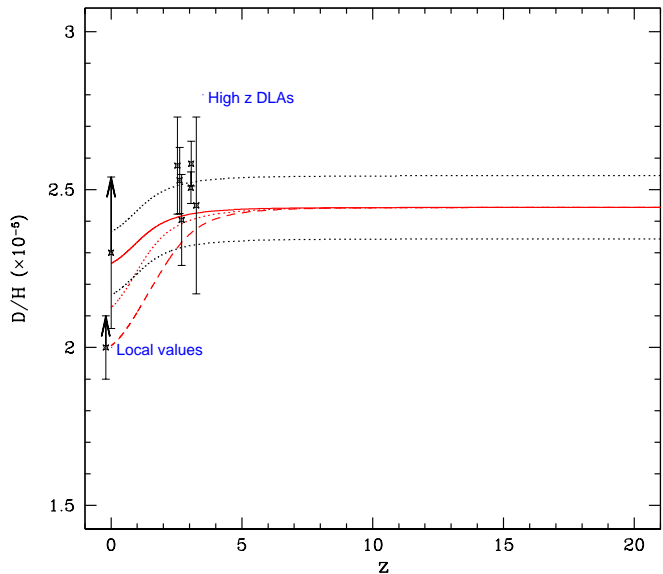} 
\caption{Cosmic deuterium evolution as a function of redshift. The deuterium evolution using the  SFR shown in Fig.~\ref{f:sfr}.  
The red solid curve corresponds to the evolution of D/H using our mean BBN value, whereas the black dotted curves correspond to 
the higher and lower (2$\sigma$) limits. High~$z$ DLAs observations 
come from Cooke \etal\ \cite{Coo14} and Riemer-S{\o}rensen \etal\ \cite{Rie15}, whereas local observations come from Linsky \etal\ \cite{Lin06} and Prodanovic \etal\ \cite{Pro10a}. 
The lower mass of the IMF is taken here as 0.1~$M_\odot$. Regarding the sensitivity to the IMF parameters, we show the impact 
of a different lower mass of the IMF (0.5~$M_\odot$, dotted red line) or adding a intermediate mass formation (between 2 and 8~$M_\odot$, dashed red line).}
\label{f:devol}
\end{center}
\end{figure}

Figure~\ref{f:devol} shows the evolution of D/H as a function of redshift, derived with the cosmic SFR shown in Figure~\ref{f:sfr} (solid black line).
Black dotted curves correspond to our 2$\sigma$  BBN limits, whereas the red solid line corresponds to the mean. 
The resulting astration factor is $f_D$=1.1. This cosmic evolution is in overall agreement with the observed values detailed above. 
Note, however, that a tension exists between the BBN value and the high redshift measurements in the sense that the latter
seem somewhat high.
However, note that, due to the extreme fragility of deuterium its potential destruction depends on many parameters of 
the star-formation history and, in particular, the IMF parameters.
We illustrate the impact of the variation of the mass lower limit of the IMF. The dotted red curve corresponds to a lower mass limit of 0.5~$M_\odot$
instead of 0.1~$M_\odot$ (solid red line).  
In this case the astration factor is $f_D$=1.15. 
Finally, we consider an extreme case by adding an intermediate mass star formation component (between 2--8~$M_\odot$) 
(dashed red line), leading to an astration factor  of  $f_D$=1.25.   
Even when considering these extreme modifications
of the IMF, the maximum variation is only 14\% which is not a large uncertainty compared to the error bars on observational data.
Recently, Prodanovic, Steigman and Fields \cite{Pro10b} have studied the deuterium evolution and its link with structure formation.
They show that a steady infall rate is required to reconcile the model with observations. 
Our cosmological model is in agreement with this result since in a hierarchical formation of structures, primordial gas is continuously
accreted into structures throughout the evolution. We also find, as these authors that 80\% of the initial interstellar gas is never processed 
within stars.

Since the paper is devoted to deuterium study we  do not consider other elements.. Obviously,
deuterium destruction can lead to \tro\ production but while we have D/H cosmological observations, we have 
only \tro/H local observations (in the galactic disk) and no constraints at high redshift.
Chiappini et al. \cite{Chi02} and Vangioni--Flam et al. \cite{Van03} have analyzed the behavior of \tro\ in the Galaxy. 
The best observational constraints come from  Bania, Rood \&  Balser \cite{Ban02}. These data are concentrated  
in the galactic disk only, i.e. at high metallicity relative to the solar value ([O/H] between -0.6 and 0.2).
Vangioni-Flam et al. have shown that it is not possible to obtain a strong constraint on the baryon density using \tro\
due to this limited range of metallicity in the sample and to the limited understanding of the chemical and stellar evolution of this isotope.

To conclude, our results are in agreement with the observations, implying that the mean abundance of deuterium has only been reduced by a factor of 1.1 to 1.25 
since its formation during BBN. 
There is, however, a tension between our BBN D/H value and the high-$z$ measurements, leaving little room for a high astration factor. 
In any case, due to the low abundance of the primordial  D/H value and the local observed constraints, the astration  factor, $f_D$, 
is less than 1.25.

\section{The lithium--deuterium anti--correlation}

\begin{figure}[htb]
\begin{center}
\includegraphics[width=.5\textwidth]{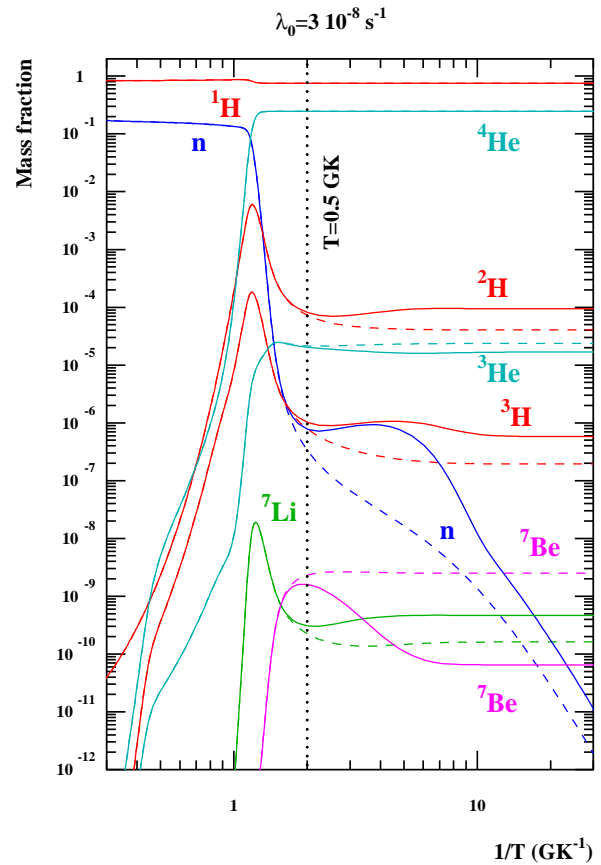}
\caption{Effect of thermal neutron injection, at a rate of $\lambda_0$=3$\times10^{-8}$~s$^{-1}$, on abundances (solid lines) as functions of the inverse of temperature (1/$T$),
compared to the standard calculation (dashed lines): $^7$Be and \tro\ abundances decrease, while \sep\ and $^3$H abundances increase.
Note the crossing of \sep\ and $^7$Be abundance curves.}
\label{f:ntemp}
\end{center}
\end{figure}

In spite of various efforts, there is still a factor of $\approx$3.5 (Table~\ref{t:heli}), between the predicted and observed
lithium primordial abundances. 
Most proposed solutions to the lithium problem lead to an increase of the deuterium production \cite{Oli12,Coc14a,Kus14}; 
they are now strongly constrained by deuterium observations. We discuss here the relation between lithium ($^7$Be+$^7$Li)
destruction and \deu\ overproduction.

At CMB deduced baryonic density, \sep\ is produced indirectly by $^3$He($\alpha,\gamma)^7$Be,
where $^7$Be will much later decay to \sep, while $^7$Be is destroyed by  $^7$Be(n,p)$^7$Li(p,$\alpha)^4$He.
The solutions to the lithium problem generally rely on an increased late time neutron abundance 
\cite{Alb12,Ren88,Jed04} to boost $^7$Be destruction  through the $^7$Be(n,p)$^7$Li(p,$\alpha)^4$He channel. 
These extra neutrons, inevitably, also boost the D and $^3$H production through the  
$^1$H(n,$\gamma)^2$H and $^3$He(n,p)$^3$H channels, respectively \cite{Kus14}. 
For instance, Fig.~\ref{f:ntemp} displays the effect of the injection of thermalized neutrons at a constant rate 
(as in Albornoz V\'asquez et al. \cite{Alb12}).

Even though other destruction channels by other thermalized \footnote{As opposed to non--thermalized particles
originating from the decay of massive relic particles during BBN (see e.g., Ref.~\cite{Cyb13}).}  
particles (p, d, t, \tro\ and $\alpha$)
have been investigated \cite{Cha11}, neutron capture remains the
only efficient one. 
Neutron induced reaction rates vary far less with temperature compared to charged-particle induced reaction rates at BBN 
temperatures.  For instance, a factor of $\sim3\times10^{-5}$ for the $^7$Be(p,$\gamma)^8$B is to be compared to a factor 2 for the
$^7$Be(n,p)$^7$Li reaction, when the temperature drops from 1.0 to 0.1~GK.
This is obviously directly linked to Coulomb barrier penetration. After $\approx$ 700~s,  
when the $^7$Be abundance has reached its maximum (Fig.~\ref{f:ntemp}), the temperature is lower than  0.5~GK.
This low temperature prevents charged particle reactions to be efficient, as it can be seen in Fig.~\ref{f:ntemp} by the flat (dashed lines) 
final evolution of the abundances. 
(In any case, the $^7$Be(p,$\gamma)^8$B reaction has such a low $Q$--value (0.1375~MeV), that the reverse reaction,
photo disintegration, is so effective that it prevents $^7$Be destruction by proton capture.)      
This could only be circumvented by the presence of strong resonances in some charged particle induced 
reactions, like $^7$Be(d,p)2$\alpha$. 
However, experiments have not supported such a nuclear physics solution involving
new conventional neutron sources \cite{Coc12b} or new resonances \cite{Kir11,Ham13} in reactions with
charged particle, suggesting non--conventional neutron sources as a solution.

\begin{figure}[htb]
\begin{center}
\includegraphics[width=.5\textwidth]{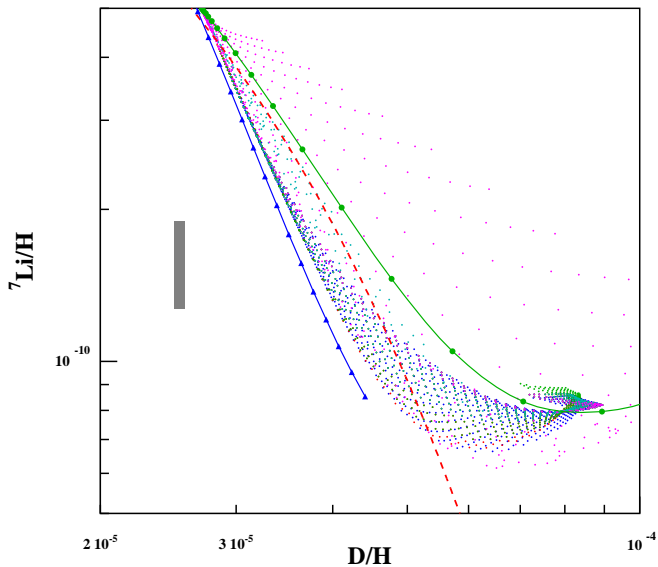}
\caption{Lithium--deuterium anti--correlation in BBN induced by different models involving neutron injection (dots: update of
Fig.~9 in Ref. \cite{Coc14a}, green circles:  Fig.~7 in Ref. \cite{Coc14a} and blue triangles, Fig.~12 in Ref. \cite{Coc12a}). 
The grey box represents the observational constraints \cite{Sbo10,Coo14}, while the
dashed line [Eq.~(\ref{q:lid4})] is a qualitative explanation of the anti--correlation.}
\label{f:scatter}
\end{center}
\end{figure}

Figure~\ref{f:scatter} is adapted from Fig. 9 in Ref.~\cite{Coc14a} summarizing the results of different models
than include late time neutron injection aiming at reducing the $^7$Be+$^7$Li production, but at the expense 
of \deu\ overproduction. These models involve mirror neutrons, dark matter decay or annihilation as 
extra neutron sources. The Figure also displays the results of a BBN calculation allowing for a coupled variation
of constants as described in Ref.~\cite{Coc12a}, where the extra neutron source arises from the
change induced in the $^1$H(n,$\gamma$)$^2$H rate.

Figure~\ref{f:ntemp} shows that $^7$Be increased destruction by neutrons is  counterbalanced by \sep\ increased production.
First, with a higher $^3$H abundance, due to the $^3$He(n,p)$^3$H increased efficiency,  the $^3$H($\alpha,\gamma)^7$Li channel, 
normally negligible at $\eta_\mathrm{CMB}$, may become dominant. 
Second, the $^7$Be increased destruction by $^7$Be(n,p)$^7$Li, produces \sep\  that is not efficiently destroyed anymore 
by $^7$Li(p,$\alpha)^4$He, because of the low temperature.
This explains that when increasing the rate of injection of
extra neutrons, the Li=$^7$Be+\sep\ abundance reaches a minimum as seen in Fig.~\ref{f:scatter} (or in
Fig.~7 of Ref.~\cite{Coc14a}).
This lower limit on Li/H, due to the transition from $^7$Be to $^7$Li direct production (Figs.~\ref{f:ntemp}--\ref{f:scatter}), 
can be compared to the minimum in Li/H as a function of $\eta$. 
 
Figure~\ref{f:ntemp} also shows that the effect of $^7$Be destruction by extra neutrons is efficient only
below $\approx$ 0.5~GK when the charged particle reactions are inefficient due to the Coulomb barrier  
and the low abundance of reactants. If we call ${\delta}Y_\mathrm{n}(t)$ the extra, late time, neutron overabundance, 
the extra destruction of $^7$Be is given by 
\begin{equation}
\frac{\mathrm{d}Y_\mathrm{^7Be}}{\mathrm{d}t}=-Y_\mathrm{^7Be}{\rho}\mathrm{N_A}\langle\sigma{v}\rangle_{\rm be7np}{\delta}Y_\mathrm{n},
\label{q:lid1}
\end{equation}
where the $Y(t)$'s are mole fractions, $\rho(t)$ the baryonic density and $\mathrm{N_A}\langle\sigma{v}\rangle$ the
thermonuclear reaction rate. We neglect the $^3$He($\alpha,\gamma)^7$Be channel at those low temperatures because of the 
Coulomb barrier (Fig.~\ref{f:ntemp}). 

At the same time, an extra deuterium production from the n(p,$\gamma$)d reaction {\em is unavoidable}, so that
\begin{equation}
\frac{\mathrm{d}Y_\mathrm{D}}{\mathrm{d}t}=+Y_\mathrm{H}\rho\mathrm{N_A}\langle\sigma{v}\rangle_{\rm pn\gamma}{\delta}Y_\mathrm{n}.
\label{q:lid2}
\end{equation}
Again, we neglect charged particle reactions and the $^3$He(n,p)$^3$H neutron drain, as we do not need to calculate ${\delta}Y_\mathrm{n}(t)$.
Putting Eqs.~(\ref{q:lid1}) and (\ref{q:lid2}) together, one obtains
\begin{equation}
\mathrm{d}\ln\left(\frac{Y_\mathrm{^7Be}}{Y_\mathrm{H}}\right)=-
{\left[\frac{\langle\sigma{v}\rangle_{\rm be7np}}{\langle\sigma{v}\rangle_{\rm pn\gamma}}\right]}
\mathrm{d}\left(\frac{Y_\mathrm{D}}{Y_\mathrm{H}}\right).
\label{q:lid3}
\end{equation}
Since, the ratio of  $^7$Be(n,p)$^7$Li to $^1$H(n,$\gamma)^2$H thermonuclear reaction rates  is almost constant
(6--8$\times10^4$ for 0.1$<T<$1~GK), one can deduce that 
\begin{equation}
\frac{\mathrm{D}}{\mathrm{H}}\approx
\left.\frac{\mathrm{D}}{\mathrm{H}}\right|_0-
\left[\ln\left(\frac{\mathrm{Li}}{\mathrm{H}}\right)-\left.\ln\left(\frac{\mathrm{Li}}{\mathrm{H}}\right)\right|_0\right]
\times1.4\times10^{-5},
\label{q:lid4}
\end{equation}
which is displayed (red dashed line) in Fig.~\ref{f:scatter}.
This is an approximation, as long as $Y_{^7\mathrm{Li}}{\ll}Y_{^7\mathrm{Be}}$, but it gives a qualitative 
explanation for the lithium--deuterium anti--correlation in most models aiming at solving the lithium overproduction. 
Depending on the precise timing of neutron injection, 
and hence, of the temperature, the efficiency of neglected reactions [e.g., d+d and $^7$Li(p,$\alpha)^4$He] need
to be considered \cite{Kus14}, but which would increase the complexity of the calculations. 
Here, we only considered thermalized neutron injection, first, because thermalization of high energy neutrons is fast \cite{Alb12}.
Second, it was already noted by Kusakabe et al. \cite{Kus14} that the ratio of $^1$H+n to $^7$Be+n cross sections increases with
energy, rendering less efficient the injection of non--thermalized neutrons for destroying $^7$Be without overproducing deuterium, when
compared with thermal neutron injection.

\section{Conclusions}

It has long been recognized that the agreement between BBN predictions and observations is quite satisfactory except for lithium.
Now that the observations of D/H in cosmological clouds, thought to be representative of the primordial abundance, 
have reached a high precision \cite{Coo14}, nuclear cross sections of all reactions leading to deuterium destruction should be 
determined with an equivalent precision \cite{DiV14}, i.e. at the percent level. To improve the precision on the three
main reaction rates governing deuterium destruction, we have re-evaluated existing experimental data, using
\sfac{s} provided by ab initio theoretical models. This is at variance with most other reaction rate evaluations
that rely on phenomenological approaches (e.g., polynomial or R--matrix) fits of experimental data. We paid 
special attention to systematic uncertainties in selection of the data sets to be considered. In particular, for 
the \ddn\ and \ddp\ \sfac{s}, we take advantage of the mostly model independent ratio of cross sections to
evaluate experimental results. The three reaction rates are found to be slightly higher than the previous
R--matrix analysis of DAACV \cite{Des04}, leading to a small but significant decrease of the D/H prediction,
$(2.45\pm0.05)\times10^{-5}$. We calculate the cosmological evolution of deuterium from BBN until present,
within a hierarchical model of structure formation and obtained a value of D/H = $(2.42\pm0.05)\times10^{-5}$,
at the redshift ($z\approx$ 3.0) of the observed cosmological clouds. 
This predicted value is compatible at the $2\sigma$ level with the observations     
$(2.53\pm0.04)\times10^{-5}$  \cite{Coo14}.

Deuterium predictions are also highly important, in relation with the lithium problem.
Most solutions involve a $^7$Be destruction by a late time neutron injection. 
We show that this is unavoidably correlated with an increase of the deuterium production by
the effect of the $^1$H(n,$\gamma)^2$H reaction.  Hence, most solutions to the lithium problem
are now severely constrained, also by deuterium precise observations.

Further progress in the domain are expected from future experiments, in particular, concerning the
\dpg\ reaction, planned to be measured at the BBN energies, at the Gran Sasso National Laboratory (LUNA),
but also, possibly, from improved theory \footnote{ 
A very recent, improved ab initio calculation of the \dpg\ \sfac\ \cite{Mar15} have lead to a reduced D/H prediction of 
$(2.49\pm0.03\pm0.03)\times10^{-5}$ compared to the previous value of $(2.61\pm0.14)\times10^{-5}$ \cite{Planck15})
.}.
Keeping systematic uncertainties on nuclear cross section measurements
at the percent level, is indeed a challenge. These systematics can be evaluated by comparing independent
measurements, with the help of theoretical \sfac{s} when data sets span different energy ranges.

\begin{acknowledgments}
This work made in the ILP LABEX (under reference ANR-10-LABX-63) was supported by French state funds managed by the ANR 
within the Investissements d'Avenir programme under reference ANR-11-IDEX-0004-02 and by the ANR VACOUL, ANR-10-BLAN-0510. 
\end{acknowledgments}

\appendix

\section{Normalization procedure}
\label{s:norm}

In other evaluations \cite{NACRE,Des04,Cyb04}, the \sfac\ shapes (polynomial, R--matrix) were fitted on experimental data. Here,
we assume that theory ($F(E)$) gives a good description of the shape but may need a scaling factor $\alpha$ (i.e. $F\to{\alpha}F$).
Calculations involve an energy dependence, and an overall normalization. The energy dependence
	is essentially provided by the Coulomb functions, and is therefore more reliable than the
	normalization, which is more sensitive to the model assumptions. Accordingly, we introduce
	a possible scaling of the theoretical calculations with a renormalization factor
	$\alpha_k$, close to unity.
By minimizing the $\chi^2$,
\begin{equation}
\chi^2(\alpha_k)=\sum_{i_k}\frac{\left[S(E_{i_k})-{\alpha_k}F(E_{i_k})\right]^2}{\sigma_{i_k}^2}
\label{q:chi2}
\end{equation}
where $S(E_{i_k})$ and $\sigma_{i_k}$ are the experimental \sfac{s} and associated uncertainties, one obtains
the scale factor best value ($\hat\alpha_k$) and associated uncertainty $(\sigma_{\hat\alpha;k})$ (for a given experiment labelled $k$ for future use) that are 
given by Eqs.~(\ref{q:alpha}) and (\ref{q:dalpha}) [Eqs. (6) and  (7) in \cite{Cyb01}]
\begin{equation}
\hat\alpha_k=\frac{\sum_i{S(E_{i_k})}F(E_{i_k})/\sigma_{i_k}^2}{\sum_{i_k}{F^2(E_{i_k})/\sigma_{i_k}^2}}
\label{q:alpha}
\end{equation}
\begin{equation}
\sigma_{\hat\alpha;k}=\frac{1}{\sqrt{\sum_{i_k}F^2(E_{i_k})/\sigma_{i_k}^2}}.
\label{q:dalpha}
\end{equation}

This takes well into account the effect of statistical uncertainties but leaves aside systematic uncertainties,
in particular on the normalization.
Systematics coming from different normalizations, from one data set to the other, 
will play an essential role. 
It is usually recommended \cite{PDG}, for incompatible data sets, to inflate
the classical error [Eq.~(\ref{q:dalpha})], by a factor of $\sqrt{\chi_\nu^2}$ so that the final reduced $\chi^2$ equals 1 
($\sqrt{\chi_\nu^2}=1)$. This method has however been questioned. We present in this section several
options that have been considered, apply them to experimental data sets in the following sections, and compare the results.

\subsection{The global data fit option} 
\label{s:ils}
 
Cyburt, Fields \& Olive \cite{Cyb01} used this procedure to re--normalize the NACRE  \sfac\ fits,
this is identical to our goal except that we use theoretical \sfac\ instead.

The value of the \nfac, $\alpha$, is again given by Eq.~(\ref{q:alpha}) [their Eq.~(6)], except that now the sum runs over all the data 
points  ($i_k=1\ldots{N_k}$) within all the $N$ data sets.

\begin{equation}
\hat\alpha=\frac{\sum_{k=1}^{N}\sum_{i_k=1}^{N_k}{S_k(E_{i_k})}F(E_{i_k})/\sigma_{i_k}^2}
{\sum_{k=1}^{N}\sum_{i_k=1}^{N_k}{F^2(E_{i_k})/\sigma_{i_k}^2}}.
\label{q:acfo}
\end{equation}
But the $\sigma_\alpha$ value proposed by Cyburt \etal\ \cite{Cyb01} is now given by their Eq.~(11) instead of their Eq.~(7) [our Eq.~(\ref{q:dalpha})]:
\begin{equation}
\sigma_{\hat\alpha}^2=\frac{\sum_{k=1}^{N}\sum_{i_k=1}^{N_k}\left[S_k(E_{i_k})-\hat\alpha F(E_{i_k})\right]^2/\sigma_{i_k}^2}
                                             {\sum_{k=1}^{N}\sum_{i_k=1}^{N_k}{\hat\alpha^2F^2(E_{i_k})/\sigma_{i_k}^2}}.
\label{q:dcfo}
\end{equation}
We note that it corresponds to  Eq.~(\ref{q:dalpha}) (with $F\to\hat\alpha F$) i.e. the classical error, but inflated by
$\sqrt{\chi^2}$ instead of $\sqrt{\chi_\nu^2}$ as it is usually recommended for incompatible data sets \cite{PDG}.

\subsection{The joint statistical and normalization fit option} 
\label{s:eux}

The method from D'Agostini \cite{DAg94} has been used, in particular,  by Serpico \etal\ \cite{Ser04} for BBN reaction rate
evaluations,  Cyburt and Davids \cite{Cyb08a}, and Sch\"urmann \etal\ \cite{Sch12},  for the \hag, and \cago, reactions respectively. 
In addition to the parameters of the theoretical model
(a single one, $\alpha$, in our case), scale factors, $\omega_k$, with associated errors, $\epsilon_k$, affect all data sets.  
Note that the $\epsilon_k$, for each experiment, are not always available. 
In that case, Serpico \etal\ \cite{Ser04} write
"{\em Whenever only a total error $\sigma_{i_k}^{tot}$ 
determination is available for a certain experiment, that error is used 
instead of $\sigma_{i_k}$, and an upper limit on the scale error is estimated as max[$\sigma_{i_k}^{tot}/Si_{i_k}$].}"
The $\chi^2$ to be minimized has the form:
\begin{equation*}
\chi^2(\alpha,\mathbf{\omega})=
\end{equation*}
\begin{equation}
\sum_{k=1}^{N}\left(\sum_{i_k=1}^{N_k}\frac{\left[\omega_kS(E_{i_k})-{\alpha}F(E_{i_k})\right]^2}{\omega_k^2\sigma_{i_k}^2}
+\frac{(\omega_k-1)^2}{\epsilon_k^2}\right)
\label{q:chi2o}
\end{equation}
Hence, the experimental values are scaled by factors that are constrained by the experimental uncertainty on
normalization while the theoretical function is also scaled by the factor we want to determine. The minimization
procedure is no longer trivial and we have to perform it numerically with the use of MINUIT  \cite{MINUIT94}. 
It is no longer possible to give an analytical expression of the uncertainty. 
The uncertainty adopted by Serpico \etal\ \cite{Ser04} is more empirical:
"{\em The overall scale error used in the analysis was chosen to be equal to the lowest experimentally determined $\epsilon_k$ for that reaction.
.....It was added in quadrature to the statistical error in the fits....}".
Cyburt and Davids \cite{Cyb08a} used the Markov Chain Monte Carlo technique to calculate the uncertainties. 
Note that in our case, the situation is simpler since, we do not fit the shape of the \sfac, which comes from theory, but
only the \nfac, and we use the error on the $\alpha$ parameter [Eq.~(\ref{q:chi2o})] provided by MINUIT \cite{MINUIT94}. 

\subsection{Our method} 
\label{s:nous}

From an other point of view, the recommended \nfac\ can be given by the weighted average of the $\hat\alpha_k$ obtained 
by Eqs.~(\ref{q:alpha}) and (\ref{q:dalpha}) from different experiments (labelled $k$)  
\begin{equation}
\bar\alpha=\sum_{k=1}^N\frac{\hat\alpha_k}{\sigma_{\hat\alpha;k}^2}\left(\sum_{k=1}^N\frac{1}{\sigma_{\hat\alpha;k}^2}\right)^{-1}.
\label{q:amean}
\end{equation}
Working out the algebra, starting from Eqs. (\ref{q:alpha}) and (\ref{q:dalpha}), one finds that this equation is
just a re--phrasing of Eq.~(\ref{q:acfo}) when no extra normalization error, $\epsilon_k$, has to be introduced.
On the contrary, it is easily introduced in Eq.~(\ref{q:amean}) by the change
\begin{equation}
\sigma_{\hat\alpha;k}^2\to\sigma_{\hat\alpha;k}^2+\epsilon_k^2.
\label{q:epsi}
\end{equation}

The error on $\bar\alpha$ would normally be given by
\begin{equation}
\sigma_{\bar\alpha}^2=\left(\sum_k\frac{1}{\sigma_{\hat\alpha;k}^2}\right)^{-1}.
\label{q:emean}
\end{equation}
We considered the possibility of applying the Cyburt \etal\ rescaling  \cite{Cyb01,Cyb04} by $\sqrt{\chi^2}$ 
to Eq.~(\ref{q:emean}),
together with an extra $N/(N-1)$, in order to obtain a weighted empirical variance
\begin{equation}
\sigma_{\bar\alpha}^2=\frac{N}{N-1}\sum_{k=1}^N\frac{(\hat\alpha_k-\bar\alpha)^2}{\sigma_{\hat\alpha;k}^2}\left(\sum_{k=1}^N\frac{1}{\sigma_{\hat\alpha;k}^2}\right)^{-1}.
\label{q:avar}
\end{equation}
(Apart from the N/N-1 factor, this is just Eq.~(21) of Cyburt \cite{Cyb04}.)
This has the advantage of converging to the empirical variance [$\sum(X_i-\bar{X})^2/(N-1)$] when the $\sigma$'s are all 
equal or favoring the contributions of the terms with lower $\sigma$'s if it is not the case.
If systematics are negligible, the central values would be the same, but Eq.~(\ref{q:dcfo}) would give a larger uncertainty.
However, we found that, {\em after introducing the normalization error with} Eq.~(\ref{q:epsi}), the reduced chi--square was
always close to unity. Accordingly, we found unnecessary to inflate the uncertainty by a  $\sqrt{\chi^2}$ factor.

\section{The D\lowercase{(p},$\gamma)^3$H\lowercase{e} data}
\label{a:dpg}

The \sfac\ is related to the total cross section by
\begin{equation}
S(E) = \sigma(E) E \exp\left(\frac{0.810799}{\sqrt{E}}\right), 
\end{equation}
(MeV and barn units).
In the following, we detail the experimental data that we considered in this evaluation, in general taken
from published tables, but when scanned from a figure, we provide here tables of the extracted numerical values. 


The Casella \etal\ (LUNA) data  \cite{Cas02} comes from their Table~I where ``{\em 
only accidental errors are reported: the systematic uncertainties ranged from 3.6\% (E$_\mathrm{eff}$ = 21.23 keV, highest measured energy) to 5.3\% 
(E$_\mathrm{eff}$  = 2.52 keV, lowest measured energy) and are negligible in comparison with the accidental errors}''. We hence adopt $\epsilon$ = 0.045
as an average systematic uncertainty.


It was found, that NACRE overlooked the overall scaling factor of 1.37 \cite{Sch96} missing in the 
Schmidt \etal\ \cite{Sch95} data. 
Here, we follow DAACV and  use instead $A_0$ (multiplied by  $4\pi$) from Table~II in  
Schmidt \etal\ \cite{Sch97}. The uncertainties reported in their Table~I are statistical only. 
The systematic error is evaluated in their \S~II.H to be $\epsilon$ = 0.09 and is used
in their Fig.~13. 
In our Table~\ref{t:sch} we present the \sfac\ used in our work.
\begin{table}[h]
\caption{Schmid \etal\ \cite{Sch97} data} 
\begin{tabular}{c|c|c}
\hline
\hline
$E_{CM}$ (MeV) & $\sigma$ (nb) & $S$ (eV.b)\\
\hline
  0.0100  &  7.30$\pm$0.38 &    0.2425 $\pm$ 0.0125 \\
  0.0167  &  30.79 $\pm$   0.84  &  0.2740  $\pm$  0.0075 \\
  0.0233 &  73.26   $\pm$ 1.38  &  0.3452 $\pm$   0.0065 \\
  0.0300  & 122.8  $\pm$  1.9  &  0.3974  $\pm$  0.0061 \\
  0.0367 &  176.0  $\pm$  2.3   & 0.4452 $\pm$   0.0057 \\
  0.0433 &  222.4  $\pm$  3.4  &  0.4738  $\pm$  0.0072 \\
  0.0500  & 252.6 $\pm$   3.4  &  0.4744 $\pm$   0.0064 \\
\hline
\hline
\end{tabular}
\label{t:sch}
\end{table}
\subsection{The Ma \etal\ data}

The \sfac, obtained from Fig.~9 in Ma \etal\ \cite{Ma97}, is shown in our Table~\ref{t:ma}.
\begin{table}[h]
\caption{Ma \etal\ \cite{Ma97} data} 
\begin{tabular}{c|c}
\hline
\hline
$E_{CM}$ (MeV)  & $S$ (eV.b)\\
\hline
0.075 &   0.685 $\pm$   0.070\\   
0.107 &   0.708  $\pm$  0.068 \\  
0.133 &   0.956  $\pm$  0.084 \\   
0.173 &   1.26  $\pm$  0.10 \\   
\hline
\hline
\end{tabular}
\label{t:ma}
\end{table}
The systematic uncertainty is estimated to be $\epsilon$ = 0.09, but may be already included in the
error bars of their Fig.~9. 


NACRE used data form W\"olfli \etal\ \cite{Wol67} (in German), presumably scanned from their
Fig.~6 or from Fig.~1 in Ref.~\cite{Wol66}  which is a comparison with theory.  
The lowest energy data point reported by NACRE at $E_{CM}$ is probably an error since
the experiment explored the $E_p$ = 2.--12. MeV range. Only three data points are found
below $E_{CM}$ = 2.~MeV, the limit of the theoretical calculation. In addition, the evaluation
of  the systematic uncertainty is difficult from the publications. 


NACRE, NACRE--II \cite{NACRE2} and Serpico \etal\ \cite{Ser04} use the data  
from Geller \etal\ \cite{Gel67}. However, as shown in their Fig.~2 \cite{Gel67}, and text,
all data are normalized to the Gunn--Irving theoretical cross section \cite{Gun51} at $E_p$ = 3.07~MeV.
Therefore, we do not use these data for normalization.


NACRE used cross--section data for \tro\ photodisintegration, scanned from Fig.~2 in Warren et al \cite{War63}, 
and applied detailed balance theorem to obtain the following \dpg\ cross section:
\begin{table}[h]
\caption{Warren et al \cite{War63} data} 
\begin{tabular}{c|c|c} 
\hline
\hline
$E_{CM}$ (MeV) & $\sigma$ ($\mu$b) & $S$ (eV.b)\\
\hline
   0.637   &  3.2 $\pm$ 0.3  &   5.7  $\pm$   0.5\\
   1.465  &  5.3 $\pm$ 0.9 &   15. $\pm$ 2. \\
   1.575  &   5.2 $\pm$ 0.6 &   16. $\pm$ 2.\\
\hline
\hline
\end{tabular}
\label{t:war}
\end{table}

\noindent
There is no information on systematic uncertainty.

NACRE used experimental data from  Griffith et al \cite{Gri62,Gri63}, presented here in Table~\ref{t:gri} . 
The low energy \sfac\ was obtained by scanning Fig.~6 in Ref.~\cite{Gri63} while the high energy
cross section is taken from Table~I in Ref.~\cite{Gri62}.
Data between parenthesis are relative measurements and are not used in the fits.

\begin{table}[ht]
\caption{Griffith et al \cite{Gri62,Gri63} data} 
\label{t:gri}
\begin{tabular}{c|c|c} 
\hline
\hline
$E_{CM}$ (MeV) & $\sigma$ ($\mu$b) & $S$ (eV.b)\\
\hline

   0.015 &    0.039 $\pm$  0.008 &    0.43  $\pm$  0.09 \\
   0.016 &    0.044  $\pm$ 0.011 &    0.42 $\pm$  0.10 \\
   0.018  &   0.052 $\pm$ 0.009 &    0.39 $\pm$   0.07 \\
   0.020 &    0.063 $\pm$  0.014 &    0.38 $\pm$   0.09\\
   0.022 &    0.077 $\pm$  0.011 &    0.39 $\pm$    0.06\\
   0.023 &    0.086 $\pm$   0.011 &    0.41 $\pm$ 0.05\\
   0.024 &    0.092 $\pm$    0.016 &    0.41 $\pm$ 0.07\\
   0.026 &    0.12  $\pm$     0.02 &    0.47 $\pm$   0.07\\
   0.027 &    0.12 $\pm$ 0.02 &    0.44 $\pm$  0.06\\
   0.028 &    0.12 $\pm$ 0.2 &    0.44 $\pm$  0.07\\
   0.031 &    0.15 $\pm$  0.02 &    0.47  $\pm$   0.06\\
   0.032 &    0.14 $\pm$ 0.02 &    0.41   $\pm$   0.07\\
\hline
   0.183   &  0.97 $\pm$    0.11   &  1.18 $\pm$    0.13\\
   0.387    & (2.20 $\pm$  0.25)  &   (3.13 $\pm$  0.36)\\
   0.503   &  2.71 $\pm$   0.13  &   4.28  $\pm$  0.21\\
   0.657   &  3.50 $\pm$  0.38   &  6.25 $\pm$    0.68\\
   1.167   &  (4.92 $\pm$   0.50)  &  (12.16   $\pm$     1.24)\\
\hline
\hline
\end{tabular}
\end{table}


DAACV quote the data from Bailey \etal\ \cite{Bai70}. Table~\ref{t:bai} displays our own scanned  data from 
Fig.~1 of that Bailey \etal\ article \cite{Bai70}.  These data have nothing in common with the presumably erroneous data 
displayed in Fig. 1.a of DAACV.
\begin{table}[h]
\caption{Bailey \etal\ \cite{Bai70} data} 
\label{t:bai}
\begin{center}
\begin{tabular}{c|c|c} 
\hline\noalign{\smallskip}
\noalign{\smallskip}\hline\noalign{\smallskip}
$E_{CM}$ (MeV) & $\sigma$ ($\mu$b) & $S$ (eV.b)\\
\hline
   0.067  & 0.43  $\pm$   0.06 &    0.66  $\pm$   0.09 \\
   0.092  & 0.67  $\pm$   0.07 &   0.88  $\pm$   0.09 \\
   0.260  & 1.55 $\pm$    0.11 &   1.98 $\pm$    0.14 \\
   0.311  & 1.77  $\pm$    0.12 &   2.36  $\pm$    0.15 \\
   0.342  & 2.01 $\pm$  0.11 &   2.75 $\pm$    0.14 \\
   0.411  & 2.26   $\pm$  0.11 &  3.29  $\pm$    0.16 \\
   0.432  & 2.34 $\pm$  0.08 &    3.47  $\pm$   0.12 \\
   0.462  & 2.46  $\pm$  0.10 &   3.75 $\pm$    0.15 \\
   0.528  & 2.69  $\pm$   0.12 &    4.34  $\pm$    0.19 \\
   0.660  & 3.31 $\pm$    0.19 &   5.93  $\pm$   0.33 \\
   0.727  & 3.43 $\pm$    0.190 &    6.45  $\pm$   0.35 \\
\noalign{\smallskip}\hline\noalign{\smallskip}
\noalign{\smallskip}\hline
\end{tabular}
\end{center}
\end{table}

NACRE--II used the data from Table~3 in Bystritsky \etal\ \cite{Bys08b} (copied in Table~\ref{t:bri}); 
the systematic uncertainty,  is less than 8\%.
\begin{table}[h]
\caption{Bystritsky al. \cite{Bys08b} data} 
\label{t:bri}
\begin{tabular}{c|c}
\hline
\hline
$E_{CM}$ (keV)  & $S$ (eV.b)\\
\hline
8.07 $\pm$ 0.27 &   0.237 $\pm$   0.071\\   
9.27 $\pm$ 0.33  &   0.277  $\pm$  0.064 \\  
9.87 $\pm$ 0.40 &   0.298  $\pm$  0.065 \\   
\hline
\hline
\end{tabular}
\end{table}
\subsection{Miscellaneous}

Unlike DAACV, we do not use the Skopik \etal\ \cite{Sko79} data  because the energies are well above
the limits of our adopted theoretical model.

\section{The D\lowercase{(d,n)}$^3$H\lowercase{e} and D\lowercase{(d,p)}$^3$H data}
\label{a:dd}

In the range of energy we are interested in, all experiments have measured both of these reaction cross sections.
This allows us to perform a test to evaluate the coherence of the data because the ratio of these cross section
is essentially governed by the Coulomb interaction, and as such is weakly dependent of the nuclear model.
We take as reference the recent ab initio calculation of  
 Arai \etal\ \cite{Ara11}. As mentioned before, the theoretical energy dependence is more accurate than the
normalization. Therefore the ratio of both cross sections is expected to be quite reliable, and
offers a good test of the various experimental data. 

\begin{figure}[htb]
\begin{center}
 \includegraphics[width=.5\textwidth]{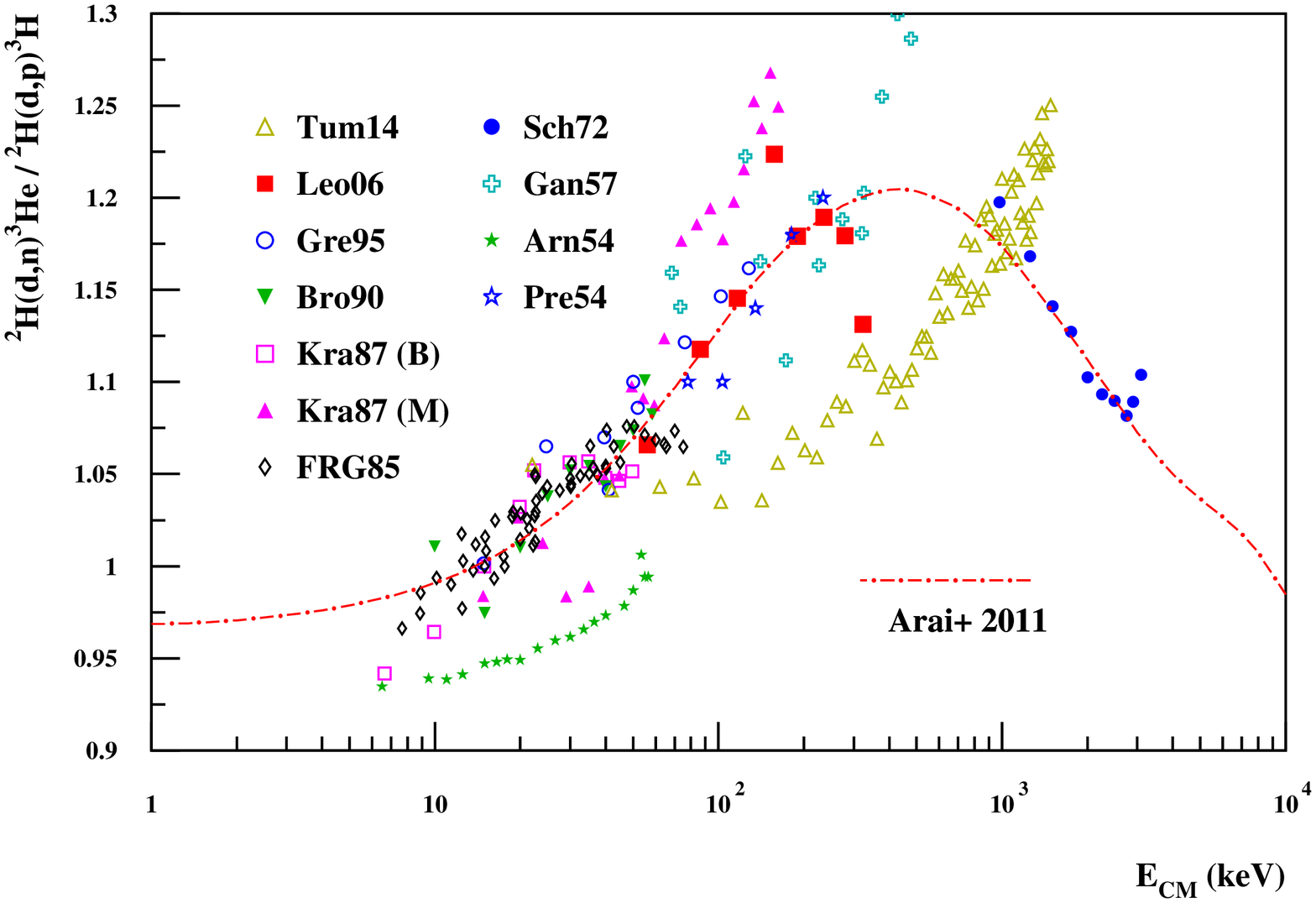} 
\caption{Ratios of \ddn\ to \ddp\ experimental and theoretical \sfac{s} from Arai \etal\ \cite{Ara11}.}
\label{f:ratio}
\end{center}
\end{figure}

The curve in Fig.~\ref{f:ratio} represents the ratio of theoretical cross 
sections compared with the data sets to be discussed below. 
Confidence in this theoretical shape is reinforced by fact that the most recent directly measured \cite{Leo06} and higher energy \cite{Sch72} experimental data sets 
follow precisely (except for a few data points) this curve. This can hardly be accidental: even though the model, because
of limitations on angular momentum, underestimates the high energy absolute cross--sections, the ratio is well reproduced,
suggesting that it is indeed independent of the nuclear matrix elements.
Obviously, this comparison is of no use to identify systematic errors in normalization which would affect both reactions in the same way.
However, deviations from the theoretical ratio may indicate normalization errors in at least one of the reaction.

The \sfac\ is related to the total cross section by:
\begin{equation}
S(E) = \sigma(E) E \exp\left(\frac{0.992857}{\sqrt{E}}\right) .
\end{equation}

In the following subsections, we detail the experimental data that we considered in this evaluation. 


NACRE and DAACV used the \sfac\ and uncertainties from Schulte \etal\ \cite{Sch72}, but the theoretical model does not reach their energy range,
and we have to put it aside. This is unfortunate since the ratio of cross sections follow the theoretical ratio (Coulomb only) in Fig.~\ref{f:ratio}, suggesting
that the normalizations are correct, or at least differ by the same constant factor.


We considered the Trojan Horse data \cite{Tum14}, but the evolution of the ratio of cross sections (Fig.~\ref{f:ratio}) is completely different from the 
theoretical one. 
We are not able to tell whether this discrepancy is experimental or is due to the theoretical model used to extract the two-body cross section
from the three-body experimental data.
However, we note that the discrepancies seen in Fig. \ref{f:ratio}, and between Figs. \ref{f:norm2n} and \ref{f:norm2p} 
are of the order of 10\%. It is not acceptable for the precision required here,
but it would be excellent, compared with other indirect methods.


The Leonard \etal\ \cite{Leo06} data are taken from their Table~III, while their Table~V provides the error matrix and quote a 
2\%$\pm$1\% scale error, and are reported here in Table~\ref{t:leo} 

\begin{table}[h]
\caption{Leonard \etal\ \cite{Leo06} data} 
\label{t:leo}
\begin{tabular}{c|c|c|c|c} 
\hline
\hline
 & \multicolumn{2}{c|}{\ddn} & \multicolumn{2}{c}{\ddp} \\ 
$E_{CM}$ & $\sigma$ & $S$   & $\sigma$ & $S$ \\
(MeV) & (mb) & (keV.b) & (mb) & (keV.b) \\
\hline
  0.0561 & 21.07 $\pm$   0.92 &   78.18 $\pm$     3.41 & 19.77 $\pm$   0.85 &   73.36 $\pm$     3.15\\
  0.0864 & 33.03 $\pm$   1.01 &   83.63 $\pm$     2.56 & 29.55 $\pm$   0.88 &   74.82 $\pm$     2.23\\
  0.1164 & 43.13 $\pm$   0.96 &   92.17 $\pm$     2.05 & 37.65 $\pm$   0.82 &   80.46 $\pm$     1.75\\
  0.1573 & 56.01 $\pm$   1.33 &  107.69 $\pm$     2.56 & 45.77 $\pm$   1.06 &   88.00 $\pm$     2.04\\
  0.1896 & 62.60 $\pm$   1.12 &  116.06 $\pm$     2.08 & 53.09 $\pm$   0.94 &   98.43 $\pm$     1.74\\
  0.2351 & 69.55 $\pm$   1.04 &  126.72 $\pm$     1.89 & 58.47 $\pm$   0.80 &  106.53 $\pm$     1.46\\
  0.2786 & 80.22 $\pm$   1.18 &  146.62 $\pm$     2.16 & 68.02 $\pm$   0.96 &  124.32 $\pm$     1.75\\
  0.3231 & 84.34 $\pm$   1.59 &  156.29 $\pm$     2.95 & 74.55 $\pm$   1.07 &  138.15 $\pm$     1.98\\
\hline
\hline
\end{tabular}
\end{table}


Table~4 in Greife \etal\ \cite{Gre95} provides \sfac{s} for the \ddp\ reaction below 15~keV, however, their Fig.~1.a shows an important
screening effect. Consequently, we do not use $E_{CM}\lesssim$15~keV data for this reaction. 
We only use the \sfac\  (Table~\ref{t:gre}) 
calculated from the cross sections found in their Table~2 \cite{Gre95} whose uncertainties  include systematics
(solid angles (3\%), gas pressure (1\%), and calorimetric measurements (1\%) [caption of Table 2]). Since they dominate over statistical errors, we adopt 
 $\epsilon$ = 0.033.

\begin{table}[h]
\caption{Greife \etal\ \cite{Gre95} data} 
\label{t:gre}
\begin{tabular}{c|c|c|c|c} 
\hline
\hline
 & \multicolumn{2}{c|}{\ddn} & \multicolumn{2}{c}{\ddp} \\ 
$E_{CM}$ & $\sigma$ & $S$   & $\sigma$ & $S$ \\
(keV) & (mb) & (keV.b) & (mb) & (keV.b) \\
\hline
     14.85   &   1.21   $\pm$     0.05  &   62.08    $\pm$    2.57 & 1.21  $\pm$      0.07   &  62.08  $\pm$      3.59 \\
     24.66    &  5.00    $\pm$    0.20   &  68.67    $\pm$    2.75 & 4.70    $\pm$    0.20   &  64.55  $\pm$      2.75\\
     39.62   &  12.60   $\pm$     0.50   &  73.21    $\pm$    2.91 & 11.80    $\pm$    0.50  &   68.56    $\pm$    2.91 \\
     41.00   &  13.20    $\pm$    0.50   &  72.92    $\pm$    2.76 & 12.70   $\pm$     0.60   &  70.16    $\pm$    3.31 \\
     50.00   &  17.90    $\pm$    0.70   &  75.89   $\pm$     2.97  & 16.30   $\pm$     0.60  &   69.10    $\pm$    2.54 \\
     52.00   &  18.10    $\pm$    0.70   &  73.21    $\pm$    2.83  & 16.70   $\pm$     0.60  &   67.55    $\pm$    2.43\\
     76.00   &  29.80    $\pm$    1.00   &  83.01    $\pm$    2.79  & 26.60   $\pm$     0.90  &   74.10   $\pm$     2.51\\
    102.00  &   39.40   $\pm$     1.40   &  89.99    $\pm$    3.20  &34.40   $\pm$     1.20  &   78.57   $\pm$     2.74 \\
    128.00  &   49.00    $\pm$    1.70  &  100.61    $\pm$    3.49   & 42.20    $\pm$    1.60  &   86.64   $\pm$     3.29\\
\hline
\hline
\end{tabular}
\end{table}


The Krauss \etal\ \cite{Kra87}  experiments took place in M\"unster ($3\lesssim{E_{CM}}\lesssim50$~keV) and at Bochum ($15\lesssim{E_{CM}}\lesssim163$~keV).
We exclude the $E<$15~keV data, because of screening (see above).
Table~2 in Krauss \etal\ \cite{Kra87} provides \sfac{s} and statistical uncertainties. A normalization error of 6.4\% comes from an 
absolute \ddp\ cross section measurement at $E_{CM}$ = 49.67~keV, to which a 5\% error due to variations in the alignment of beam and jet target profiles has to be added 
for the the M\"unster data.
NACRE added quadratically all these errors. Hence, we use $\epsilon$ = 0.064 for the Bochum data  and, following
the authors, $\epsilon$ = 0.082 for the  M\"unster data.

NACRE used the \sfac\ and uncertainties (0.4--4\%) from Brown \& Jarmie 1990 \cite{Bro90} Table~II. {\em However, NACRE did not 
take into account the 1.3\% "scale error", dominated by the uncertainty in the p +d elastic calibration} leading to $\epsilon$ = 0.013.

The article from the ``The First Research Group, The First Research Division'' \cite{first}, written in chinese, reports on
 the \ddp\ and \ddn\ cross section data, from $E_d$ =  15 to 150~keV, which are available in \cite{exfor} and have been used
 in Refs.~\cite{Ser04,Tum14}. 
The ratio of cross sections, shown in Fig.~\ref{f:ratio}, follows closely and scatters evenly around the theoretical curve. 
However, because of our inability to understand the core of the article, and in particular the error budget, we considered 
these results in our evaluation, but did not use them in our fit.

Preston \etal\ \cite{Pre54}, measured the  \ddp\ and \ddn\ cross sections, from $E_d$ =  156 to 466~keV. The ratio
of cross section is in good agreement with theory (Fig.~\ref{f:ratio}).


We list here data which may have been used in other evaluations but that we put aside in our evaluation.
Bystritsky \etal\  \cite{Bys08a} quoted in NACRE-II are not considered as they explore a range of energy
where screening is important; its study being the goal.
The same ($E<10$~keV) applies to Belov \etal\ \cite{Bel90} data obtained from \cite{exfor}.
Hofstee \etal\ (a conference proceeding) \cite{Hof01}, also quoted in NACRE-II 
is not considered either (two data points with large ($\pm5\%\pm2\%$) uncertainty). 
Davidenko \etal\  \cite{Dav57} quote a 20\% uncertainty on \ddn.
Data from Ganeev \etal\  \cite{Gan57}  can be  obtained from \cite{exfor}.
There is an overlap in energy range for the \ddn\ and \ddp\ cross section measurements but the energy steps are different.
Hence, we plotted in Fig.~\ref{f:ratio} the ratio between $\sigma_n$ {\em interpolated} experimental values  and  experimental $\sigma_p$ values.
The resulting values show a large scatter ($\approx$8\%) w.r.t. the theoretical curve. 
In McNeill and Keyser \cite{McN51}, it is stated (p. 605) that {\em ``In addition, errors in calibration and measurement, amounting to a maximum 
possible value of 20 percent, must be considered.''}, we do not include these data in our fit.  
Arnold \etal\  \cite{Arn54}, provide \ddn\ and \ddp\ cross section data from $E_d$ = 13. to 113.~keV with a detailed error analysis. 
Unfortunately, as can be seen in Fig.~\ref{f:ratio}, their $\sigma_n/\sigma_p$ ratio is too small by $\approx$7\%. This is apparently due
to an systematic error in the \ddn\ data (Fig.~\ref{f:ddn}).

\section{Tabulated reaction rates}

\begin{table*}[htb]
\caption{Present recommended D\lowercase{(p},$\gamma)^3$H\lowercase{e}, D\lowercase{(d,n)}$^3$H\lowercase{e} and D\lowercase{(d,p)}$^3$H
 rates} 
\begin{longtable}{r|l|c|l|c|l|c}
\hline
\hline
 & \multicolumn{2}{c|}{\dpg} & \multicolumn{2}{c|}{\ddn} & \multicolumn{2}{c|}{\ddp}       \\
T (GK)  & rec. rate & $f.u.$ & rec. rate & $f.u.$ & rec. rate & $f.u.$ \\
\hline
    0.001 & 4.815$\times10^{-14}$  &    1.038 & 1.142$\times10^{ -8}$  &   1.011 & 1.173$\times10^{ -8}$  &    1.011 \\
    0.002 & 6.409$\times10^{ -9}$  &    1.038 & 5.470$\times10^{ -5}$  &   1.011 & 5.609$\times10^{ -5}$  &    1.011 \\
    0.003 & 4.525$\times10^{ -7}$  &    1.038 & 3.021$\times10^{ -3}$  &   1.011 & 3.092$\times10^{ -3}$  &    1.011 \\
    0.004 & 4.896$\times10^{ -6}$  &    1.038 & 3.732$\times10^{ -2}$  &   1.011 & 3.814$\times10^{ -2}$  &    1.011 \\
    0.005 & 2.444$\times10^{ -5}$  &    1.038 & 2.212$\times10^{ -1}$  &   1.011 & 2.257$\times10^{ -1}$  &    1.011 \\
    0.006 & 8.086$\times10^{ -5}$  &    1.038 & 8.546$\times10^{ -1}$  &   1.011 & 8.707$\times10^{ -1}$  &    1.011 \\
    0.007 & 2.078$\times10^{ -4}$  &    1.038 & 2.505$\times10^{  0}$  &   1.011 & 2.549$\times10^{  0}$  &    1.011 \\
    0.008 & 4.499$\times10^{ -4}$  &    1.038 & 6.066$\times10^{  0}$  &   1.011 & 6.164$\times10^{  0}$  &    1.011 \\
    0.009 & 8.619$\times10^{ -4}$  &    1.038 & 1.278$\times10^{  1}$  &   1.011 & 1.297$\times10^{  1}$  &    1.011 \\
    0.010 & 1.506$\times10^{ -3}$  &    1.038 & 2.424$\times10^{  1}$  &   1.011 & 2.458$\times10^{  1}$  &    1.011 \\
    0.011 & 2.450$\times10^{ -3}$  &    1.038 & 4.237$\times10^{  1}$  &   1.011 & 4.290$\times10^{  1}$  &    1.011 \\
    0.012 & 3.767$\times10^{ -3}$  &    1.038 & 6.936$\times10^{  1}$  &   1.011 & 7.016$\times10^{  1}$  &    1.011 \\
    0.013 & 5.531$\times10^{ -3}$  &    1.038 & 1.077$\times10^{  2}$  &   1.011 & 1.088$\times10^{  2}$  &    1.011 \\
    0.014 & 7.816$\times10^{ -3}$  &    1.038 & 1.600$\times10^{  2}$  &   1.011 & 1.615$\times10^{  2}$  &    1.011 \\
    0.015 & 1.070$\times10^{ -2}$  &    1.038 & 2.291$\times10^{  2}$  &   1.011 & 2.310$\times10^{  2}$  &    1.011 \\
    0.016 & 1.425$\times10^{ -2}$  &    1.038 & 3.179$\times10^{  2}$  &   1.011 & 3.202$\times10^{  2}$  &    1.011 \\
    0.018 & 2.366$\times10^{ -2}$  &    1.038 & 5.667$\times10^{  2}$  &   1.011 & 5.698$\times10^{  2}$  &    1.011 \\
    0.020 & 3.659$\times10^{ -2}$  &    1.038 & 9.310$\times10^{  2}$  &   1.011 & 9.343$\times10^{  2}$  &    1.011 \\
    0.025 & 8.753$\times10^{ -2}$  &    1.038 & 2.504$\times10^{  3}$  &   1.011 & 2.502$\times10^{  3}$  &    1.011 \\
    0.030 & 1.701$\times10^{ -1}$  &    1.038 & 5.301$\times10^{  3}$  &   1.011 & 5.276$\times10^{  3}$  &    1.011 \\
    0.040 & 4.476$\times10^{ -1}$  &    1.038 & 1.568$\times10^{  4}$  &   1.011 & 1.549$\times10^{  4}$  &    1.011 \\
    0.050 & 8.915$\times10^{ -1}$  &    1.038 & 3.369$\times10^{  4}$  &   1.011 & 3.307$\times10^{  4}$  &    1.011 \\
    0.060 & 1.510$\times10^{  0}$  &    1.038 & 6.013$\times10^{  4}$  &   1.011 & 5.868$\times10^{  4}$  &    1.011 \\
    0.070 & 2.302$\times10^{  0}$  &    1.038 & 9.527$\times10^{  4}$  &   1.011 & 9.246$\times10^{  4}$  &    1.011 \\
    0.080 & 3.265$\times10^{  0}$  &    1.038 & 1.390$\times10^{  5}$  &   1.011 & 1.343$\times10^{  5}$  &    1.011 \\
    0.090 & 4.392$\times10^{  0}$  &    1.038 & 1.912$\times10^{  5}$  &   1.011 & 1.837$\times10^{  5}$  &    1.011 \\
    0.100 & 5.676$\times10^{  0}$  &    1.038 & 2.513$\times10^{  5}$  &   1.011 & 2.404$\times10^{  5}$  &    1.011 \\
    0.110 & 7.109$\times10^{  0}$  &    1.038 & 3.190$\times10^{  5}$  &   1.011 & 3.039$\times10^{  5}$  &    1.011 \\
    0.120 & 8.685$\times10^{  0}$  &    1.038 & 3.938$\times10^{  5}$  &   1.011 & 3.737$\times10^{  5}$  &    1.011 \\
    0.130 & 1.040$\times10^{  1}$  &    1.038 & 4.753$\times10^{  5}$  &   1.011 & 4.493$\times10^{  5}$  &    1.011 \\
    0.140 & 1.224$\times10^{  1}$  &    1.038 & 5.631$\times10^{  5}$  &   1.011 & 5.304$\times10^{  5}$  &    1.011 \\
    0.150 & 1.420$\times10^{  1}$  &    1.038 & 6.568$\times10^{  5}$  &   1.011 & 6.165$\times10^{  5}$  &    1.011 \\
    0.160 & 1.628$\times10^{  1}$  &    1.038 & 7.559$\times10^{  5}$  &   1.011 & 7.072$\times10^{  5}$  &    1.011 \\
    0.180 & 2.076$\times10^{  1}$  &    1.038 & 9.691$\times10^{  5}$  &   1.011 & 9.011$\times10^{  5}$  &    1.011 \\
    0.200 & 2.565$\times10^{  1}$  &    1.038 & 1.200$\times10^{  6}$  &   1.011 & 1.110$\times10^{  6}$  &    1.011 \\
    0.250 & 3.941$\times10^{  1}$  &    1.038 & 1.842$\times10^{  6}$  &   1.011 & 1.682$\times10^{  6}$  &    1.011 \\
    0.300 & 5.505$\times10^{  1}$  &    1.038 & 2.555$\times10^{  6}$  &   1.011 & 2.309$\times10^{  6}$  &    1.011 \\
    0.350 & 7.225$\times10^{  1}$  &    1.038 & 3.318$\times10^{  6}$  &   1.011 & 2.974$\times10^{  6}$  &    1.011 \\
    0.400 & 9.076$\times10^{  1}$  &    1.038 & 4.119$\times10^{  6}$  &   1.011 & 3.663$\times10^{  6}$  &    1.011 \\
    0.450 & 1.104$\times10^{  2}$  &    1.038 & 4.946$\times10^{  6}$  &   1.011 & 4.371$\times10^{  6}$  &    1.011 \\
    0.500 & 1.310$\times10^{  2}$  &    1.038 & 5.792$\times10^{  6}$  &   1.011 & 5.089$\times10^{  6}$  &    1.011 \\
    0.600 & 1.748$\times10^{  2}$  &    1.038 & 7.517$\times10^{  6}$  &   1.011 & 6.543$\times10^{  6}$  &    1.011 \\
    0.700 & 2.212$\times10^{  2}$  &    1.038 & 9.260$\times10^{  6}$  &   1.011 & 8.001$\times10^{  6}$  &    1.011 \\
    0.800 & 2.700$\times10^{  2}$  &    1.038 & 1.100$\times10^{  7}$  &   1.011 & 9.448$\times10^{  6}$  &    1.011 \\
    0.900 & 3.207$\times10^{  2}$  &    1.038 & 1.272$\times10^{  7}$  &   1.011 & 1.087$\times10^{  7}$  &    1.011 \\
    1.000 & 3.729$\times10^{  2}$  &    1.038 & 1.442$\times10^{  7}$  &   1.011 & 1.228$\times10^{  7}$  &    1.011 \\
    1.250 & 5.093$\times10^{  2}$  &    1.038 & 1.850$\times10^{  7}$  &   1.011 & 1.565$\times10^{  7}$  &    1.011 \\
    1.500 & 6.522$\times10^{  2}$  &    1.038 & 2.235$\times10^{  7}$  &   1.011 & 1.882$\times10^{  7}$  &    1.011 \\
    1.750 & 8.000$\times10^{  2}$  &    1.038 & 2.595$\times10^{  7}$  &   1.012 & 2.181$\times10^{  7}$  &    1.012 \\
    2.000 & 9.517$\times10^{  2}$  &    1.038 & 2.932$\times10^{  7}$  &   1.012 & 2.461$\times10^{  7}$  &    1.012 \\
    2.500 & 1.265$\times10^{  3}$  &    1.038 & 3.546$\times10^{  7}$  &   1.013 & 2.976$\times10^{  7}$  &    1.013 \\
    3.000 & 1.587$\times10^{  3}$  &    1.038 & 4.093$\times10^{  7}$  &   1.014 & 3.440$\times10^{  7}$  &    1.014 \\
    3.500 & 1.914$\times10^{  3}$  &    1.038 & 4.585$\times10^{  7}$  &   1.014 & 3.863$\times10^{  7}$  &    1.014 \\
    4.000 & 2.244$\times10^{  3}$  &    1.039 & 5.031$\times10^{  7}$  &   1.015 & 4.251$\times10^{  7}$  &    1.015 \\
    5.000 & 2.905$\times10^{  3}$  &    1.040 & 5.816$\times10^{  7}$  &   1.016 & 4.946$\times10^{  7}$  &    1.016 \\
    6.000 & 3.557$\times10^{  3}$  &    1.042 & 6.488$\times10^{  7}$  &   1.017 & 5.552$\times10^{  7}$  &    1.017 \\
    7.000 & 4.194$\times10^{  3}$  &    1.044 & 7.072$\times10^{  7}$  &   1.018 & 6.077$\times10^{  7}$  &    1.018 \\
    8.000 & 4.812$\times10^{  3}$  &    1.046 & 7.583$\times10^{  7}$  &   1.018 & 6.529$\times10^{  7}$  &    1.018 \\
    9.000 & 5.410$\times10^{  3}$  &    1.047 & 8.037$\times10^{  7}$  &   1.018 & 6.912$\times10^{  7}$  &    1.018 \\
   10.000 & 5.988$\times10^{  3}$  &    1.049 & 8.437$\times10^{  7}$  &   1.018 & 7.228$\times10^{  7}$  &    1.019 \\
\hline
\hline
\end{longtable}
\label{t:rates}
$f.u.$ = uncertainty factor, see Eq.~\ref{q:fu} and Ref.~\cite{Eval1}
\end{table*}

\newpage


\end{document}